\documentclass[12pt]{article}
\usepackage{epsf,amsfonts}
\newcommand{\newsection}{    
\setcounter{equation}{0}\section}
\renewcommand{\appendix}[1]{
     \addtocounter{section}{1}
     \setcounter{equation}{0}
     \renewcommand{\thesection}{\Alph{section}}
     \section*{Appendix \thesection\protect\indent #1}
     \addcontentsline{toc}{section}{Appendix \thesection\ \ \ #1}
}
\newcommand\encadremath[1]{\vbox{\hrule\hbox{\vrule\kern8pt
\vbox{\kern8pt \hbox{$\displaystyle #1$}\kern8pt}
\kern8pt\vrule}\hrule}} \def\enca#1{\vbox{\hrule\hbox{
\vrule\kern8pt\vbox{\kern8pt \hbox{$\displaystyle #1$} \kern8pt}
\kern8pt\vrule}\hrule}}
\newcommand{\rf}[1]{(\ref{#1})}
\newcommand{\eq}[1]{Eq.~(\ref{#1})}
\renewcommand{\and}{{\qquad {\rm and} \qquad}}
\newcommand{\where}{{\qquad {\rm where} \qquad}}
\newcommand{\with}{{\qquad {\rm with} \qquad}}
\newcommand{\for}{{\qquad {\rm for} \qquad}}
\newcommand{\virg}{{\qquad , \qquad}}
 \newcommand{\Tr}{{\,\rm Tr}\:}
\newcommand{\tr}{{\,\rm tr}\:}
\renewcommand{\l}{\lambda}
\newcommand{\om}{\omega}
\newcommand{\calP}{{\cal P}}
\newcommand{\ii}{{\mathrm{i}}}
\newcommand{\e}{{\,\rm e}\,}
\newcommand{\ee}[1]{{{\rm e}^{#1}}}
\renewcommand{\d}{{{\partial}}}
\newcommand{\D}{{{\hbox{d}}}}
\newcommand{\dmat}[2]{\mathrm{d}_{\scriptscriptstyle{#1}}[#2]}
\newcommand{\Pint}{{\int\kern -1.em -\kern-.25em}}
\newcommand{\Vol}{\mathrm{Vol}}
\renewcommand{\Re}{{\mathrm{Re}}}
\renewcommand{\Im}{{\mathrm{Im}}}
\newcommand{\sn}{{\rm sn}}
\newcommand{\cn}{{\rm cn}}
\newcommand{\dn}{{\rm dn}}
\newcommand{\ssq}[1]{{\sqrt{\sigma({#1})}}}
\begin{document}
\topmargin 0pt
\oddsidemargin 5mm
\headheight 0pt
\headsep 0pt
\topskip 9mm
%
%
\pagestyle{empty}
\hfill SPT-00/36
\addtolength{\baselineskip}{0.20\baselineskip}
\begin{center}
\vspace{26pt}
{\large \bf {Breakdown of universality in multi-cut matrix models }}
\newline
\vspace{26pt}

{\sl G. \ Bonnet}\hspace*{0.05cm}\footnote{
E-mail: gabonnet@spht.saclay.cea.fr
},
{\sl F. \ David}\hspace*{0.05cm}\footnote{
E-mail: david@spht.saclay.cea.fr
}\,\footnote{Physique Th\'eorique CNRS},
{\sl B.\
Eynard}\hspace*{0.05cm}\footnote{ E-mail: eynard@spht.saclay.cea.fr
}\\
\vspace{6pt}
Service de Physique Th\'{e}orique de Saclay,\\
F-91191 Gif-sur-Yvette Cedex, France.\\
\end{center}
\vspace{20pt}
\begin{center}
{\bf Abstract}
\end{center}
%
We solve the puzzle of the disagreement between orthogonal polynomials
methods and mean field calculations for random $N\times N$ matrices
with a disconnected eigenvalue support.
We show that the difference does not stem from a $\mathbb{Z}_2$
symmetry breaking, but from the discreteness of the number of
eigenvalues.  This leads to additional terms (quasiperiodic in $N$)
which must be added to the naive mean field expressions.  Our result
invalidates the existence of a smooth topological large $N$ expansion
and some postulated universality properties of correlators.  We derive the
large $N$ expansion of the free energy for the general 2-cut case.
From it we rederive by a direct and easy mean-field-like method the
2-point correlators and the asymptotic orthogonal polynomials.  We
extend our results to any number of cuts and to non-real potentials.

\newpage
\pagestyle{plain}
\setcounter{page}{1}

\newsection{Introduction}
\label{intro}

Random Matrix Models have been introduced in order to give an approximate
statistical description of quantum systems involving disorder, chaos,
complexity or whatever prevents from solving the equations of motion exactly.
Those models are described by a matrix (Hamiltonian, transfer matrix, or
scattering matrix) of large size $N$, which is too complicated to be
diagonalized exactly, and for which only statistical observations of the
spectrum are available (see \cite{reviewRMT,Mehta} for a review on RMT).

Most of the quantities of interest and observables are related to the short
range (in energy scale) behavior of the spectrum (indeed small energies
correspond to long time evolution, i.e. to equilibrium thermodynamical
properties).
This is why the short range correlation functions are the most studied.

At first, the simplest models assumed a gaussian weight for the random matrix
and gave good agreement with observations, provided that the ensemble of
matrices(hermitian, orthogonal, quaternionic...) has the required symmetries
(time reversibility,...) \cite{Mehta,Wigner}.

It has been observed that the correlation functions of the spectrum possess
universal properties at sort range, which do not depend on the
probability weight, gaussian or not.  This universality has been
proved for a wide range of models by several approaches
\cite{Wigner,rconject,univtests,rCorreldeu,BrezinZee}, but a very
general proof and the exact hypothesis which lead to it are still
under investigations.

Moreover, the long range correlation functions appear to share also
some universal properties, which depend on the probability weight only
through a few parameters \cite{AmJuMa90,BrezinZee}.
The most striking example is the 2-point
correlation function that we shall discuss below.

\bigskip

In the following we will restrict our attention to the so called
Hermitian-One-Matrix-Model\footnote{ The other ensembles are of course
worth considering, but  this one is the simplest.} \cite{Mehta,BIPZ}
(hermiticity corresponds to a
system with broken time reversibility, for instance in the presence of a
magnetic field).

We consider a hermitian matrix $M$ of size $N\times N$ with a probability
law of the form:
$$ \calP(M) = \ee{-N\tr V(M)} $$
where $V$ is a polynomial potential bounded from below.

We wish to study the statistical properties of the eigenvalues
$(\l_1,\dots ,\l_{N})$ of $M$ in the large $N$ limit, in particular the density
of eigenvalues $\rho(\l)$, and the correlation function $R(\l,\mu)$, which
measures the probability that two of the eigenvalues take the values
$\l$ and $\mu$.

Roughly speaking, the eigenvalues tend to occupy a finite interval centered
around the bottom of the potential well, and in the large $N$ limit, the
density of eigenvalues $\rho(\l)$ is a continuous function with a compact
support.

The simplest case, where the support is connected, known as the ``1-cut case'',
has been extensively studied \cite{Mehta}.
It is then found that the density $\rho(\l)$ is not universal, it
depends on the
details of the potential $V$, while the connected correlation function
$R_c(\l,\mu)$ is universal in the short range regime ($|\l-\mu|\sim O(1/N)$),
but also in the long range regime ($|\l-\mu|\sim O(1)$) once the short range
oscillations (of period $\sim O(1/N)$) have been smoothed out.

\bigskip

What happens when the potential $V$ possesses several wells, of
approximatively the same depths?
Then the density $\rho(\l)$ has a disconnected support, $[a_1,b_1] \cup
[a_2,b_2]\cup \dots \cup [a_s,b_s]$, each interval $[a_i,b_i]$ being centered
around one well of the potential $V$.
This case is known as the multicut (or multiband) case (here $s$ cuts).

In the multicut case the density is still not universal, whereas the 2-point
correlation function is universal in the short range regime and seems to have
some universal properties in the long range regime after smoothing: in
\cite{AmAk} an explicit form of the 2-point connected
correlation function was given, and is claimed to be universal:
indeed, according to the authors of \cite{AmAk} ``it depends only on
the number of connected components of the support and on the position
of the endpoints, but not on the potential''.
However, more recently several authors
\cite{kanz98,Deo,BrezinDeo} have studied the two-cut case
$s=2$: they concentrated on the case of an even potential $V$ (the two
cuts are thus symmetric $[a,b]$ and $[-b,-a]$).  Using an ansatz for
the asymptotic expression of orthogonal polynomials in the large $N$
limit, and rederiving the two-point function from this ansatz,
they observed that the connected
correlation function is still universal in the short distance regime
(which was expected), but more surprisingly, that the smoothed connected
correlation function in the long range regime depends on the parity of
$N$ ($N$ being the size of the matrix).  This seems to contradict the former
result of \cite{AmAk} !

\bigskip
In this
paper, we will solve this paradox.

\medskip

We will show that the semi-classical method of \cite{AmAk}
gives the
2-point connected correlation function only up to an additional
non-universal term, which is already present in the free energy, but
subdominant at large $N$ in this case.
We correct the semi-classical
argument of \cite{AmAk}, and give a simple (and physically appealing)
derivation of the origin of the additional term, that we compute
explicitly.
This allows us to recover the results of
\cite{kanz98,BrezinDeo} for the symmetric case, and to generalize them
to non symmetric potentials, without using orthogonal polynomials.

Using the same
semi-classical argument, we are able to derive large $N$ asymptotics for the
orthogonal polynomials, recovering the results of
\cite{kanz98,BrezinDeo} as well as the general $s$ cuts asymptotics
which appeared recently in the mathematical literature \cite{deift},
and to extend these results to the case of complex potentials.

\medskip
The effect leading to the new term in the semi-classical calculation
is simple enough to be explained briefly in this introductory section.

In \cite{AmAk}, the free energy $F$ of the matrix model is derived by a
saddle point approximation.  In particular, one has to extremize the
action with respect to variations of the number $n_i$ ($i=1\dots s$)
of eigenvalues in each connected part of the support, or in other
words with respect to the occupation ratio $x_i=n_i/N$:
\begin{equation}
     F=F(x_c)
     \qquad {\rm where}\qquad \left.  {\partial F\over \partial
     x}\right|_{x=x_c} = 0
     \label{eq:Fintro}
\end{equation}
However, one has here missed the crucial fact that $n_i=N x_i$ are not real
numbers but integers.  When $Nx_c$ is not an integer, the extremum of
$F(x)$ is never reached, and the saddle point approximation has to be
slightly modified.  Roughly speaking the discrete sum cannot be
approximated by an integral:
\begin{equation}
     \sum_{n} \ee{-g (n-Nx_{c})^2} \neq \int \D{x}\, \ee{-N^2 g (x-x_c)^2}
     \label{eq:sumintro}
\end{equation}
The discrete sum actually depends on how far from an integer $Nx_c$ is.
For instance, in the symmetric case, we have $x_c={1\over 2}$, and the result
depends on the parity of $N$.

We will show that (as expected from \eq{eq:sumintro}) in general the
result involves elliptic theta functions depending on $Nx_{c}$, thus
leading to a quasi-periodic dependence on $N$.  This effect is of
order $N^{-2}$ for the free energy, but is of order $1$ for the
computation of the orthogonal polynomials and for the correlation
functions.  It implies in particular that there is {ñ\bf no regular large $N$
topological expansion} (involving only power series in $N^{-2}$) for
the 2-cut matrix model.


We will find out that the short range correlation function is
universal, while the long range smoothed correlation depends on $N$
quasi-periodically.

\bigskip

\noindent The paper is divided as follows:
in section 2 we introduce the method and notations for the 2-cut model,
and we compute the free energy.
In section 3 we derive the 2-point correlation function, and we recover the
expression of \cite{kanz98,BrezinDeo,deift} in the symmetric case.  In section
4, we give an asymptotic expression for the orthogonal polynomials,
which we use to rederive the universal short range properties of the
spectrum, as well as the smoothed long range 2-point correlation
function.

The generalizations to a complex potential or to an arbitrary number of cuts
are presented in Appendix B and C.
Appendix A is a summary of some relationships between elliptical functions in
case the reader is not familiar with them.


%
%
\newsection{The free energy}
\label{sec.2}
\subsection{Basics}
\label{subsec.2.1}

We start from the standard Hermitian matrix model defined by the
partition function
\begin{equation}
     Z[V;N]\ =\ \int \dmat{N}{M}\,\, \ee{-N\tr V(M)}
     \label{eq:Z}
\end{equation}
where $N$ is the dimension of the matrix $M$,
$V$ is an analytic -- in general polynomial -- and for the moment
  real -- function, and $\D_{N}{[M]}$ is the standard $U(N)$ invariant
  measure over Hermitian matrices
\begin{equation}
\dmat{N}{M}\ =\ \prod_{i=1}^{N}\D M_{ii}\, \prod_{1\le i<j\le N}^{} 2\,
     \D \Re (M_{ij})\,\D\Im (M_{ij})
     \label{eq:ZV}
\end{equation}
Integrating out the "angular part" of $M$, $Z$ can be rewritten as an integral
over the $N$ eigenvalues $\l_1 ,\dots ,\l_N$ of $M$
\cite{Mehta}
\begin{equation}
     Z[V;N]\  =\  \mathbf{C}_{N}\,\tilde Z[V;N]
\label{eq:Ztilde}
\end{equation}
\begin{equation}
     \tilde Z[V;N]\ =\
     \int \prod_{k=1}^{N} \D{\l_k} \,\, \ee{-N \sum_k V(\l_k)}\,\,
     \prod_{k<l} (\l_k-\l_l)^2
     \ =\
     \int \prod_{k=1}^{N} \D{\l_k} \,\, \ee{-N^2 S(\l_1,\dots,\l_N)}
\label{eq:DM}
\end{equation}
with the measure factor
\begin{equation}
     C_{N}\ =\ \Vol\left[{\mathrm{U}(N)\over \mathrm{U}(1)^{N}
     \times{\mathfrak S}_{N}}\right]\
     =\ {1\over N!}\,\prod_{K=1}^{N}\,{(2\pi)^{K-1}\over\Gamma(K)}
     \label{eq:evZ}
\end{equation}
and with the action $S(\l_k)$:
\begin{equation}
     S(\l_1,\dots,\l_N) = {1\over N}\sum_{k=1}^N V(\l_k) - {1\over 2\,N^2}
\sum_{1\leq k\neq l\leq N} \ln{(\l_k-\l_l)^2}
     \label{eq:Sdiscr}
\end{equation}
In the simplest "one cut" case, corresponding in particular to a concave
potential, it is known \cite{BIPZ}
that the free energy $F$ defined as
\footnote{ $Z$ is normalized here so that $F$ is zero for the Gaussian
model $V={1 \over 2} M^2$}
\begin{equation}
      Z[V;N]\ =\ \left({2\pi\over N}\right)^{{N^2\over 2}}
      \,\ee{-F[V;N]}
     \label{eq:defF}
\end{equation}
has a topological large $N$ expansion
\begin{equation}
     F\ = \ N^{2}\ F_{0}\,+\,F_{1}+N^{-2}\,F_{2}\,+\,\ldots
     \label{eq:topexp}
\end{equation}
obtained for instance by re-organizing the perturbative expansion
according to the topology of the Feynman diagrams.
The large $N$ limit (planar limit) can be described by a "master field"
configuration where the eigenvalues are described by a continuous
density $\rho(\l)$ with a connected compact support ${\cal C}=[a,b]$,
with the constraints
\begin{equation}
    \int_{{\cal C}}\D \l \,\rho(\l)\ =\ 1
    \qquad\and\qquad \rho(\l)\ge 0\ \mathrm{if}\ \l\in{\cal C}
     \label{eq:constraint}
\end{equation}
and the action \ref{eq:Sdiscr} becomes
\begin{equation}
     S[\rho]= \int_{\cal C} \D\l V(\l)\rho(\l) -
\int_{{\cal C}\times{\cal
C}} \D\l \D\mu\,\, \rho(\l)\rho(\mu)\ln{|\l-\mu|}
     \label{eq:Srho}
\end{equation}
The leading term of the free energy $F_{0}$ can be obtained by the saddle
point method: the effective action $S[\rho]$ is extremized for
a continuous distribution $\rho_{c}$ and we have simply
\begin{equation}
F_{0}\ =\ S[\rho_c]
     \label{eq:spaction}
\end{equation}(up to an additive -- potential independent -- constant).
To compute $\rho_{c}$ we include the constraint \ref{eq:constraint} in the
effective action by a Lagrange multiplier $\Gamma$
\begin{equation}
      {\bar S}[\rho]\ =\ S[\rho]\,+\,\Gamma(1-\int_{{\cal C}}\rho)
     \label{eq:barS}
\end{equation}
The saddle point equation for $\rho$ reads:
\begin{equation}
     {\d {\bar S}\over \d\rho(\l)}
\ = \ V(\l) - 2\,\int_{\cal C} \D\mu\,
\rho(\mu)\,{\ln(|\l-\mu|)}\,-\,\Gamma\ =\ 0 \qquad \forall\, \l\in{\cal
C}
     \label{eq:spequ}
\end{equation}
which simply means that the real part of the effective potential
\begin{equation}
     V_{\rm eff}(\l) = V(\l) - 2\,\int_{\cal C} \D\mu\,\rho(\mu)\,\ln{(\l-\mu)}
     \label{eq:Veff}
\end{equation}
is constant on the e.v. support ${\cal C}$, and equal to $\Gamma$.
The derivative of \ref{eq:spequ} w.r.t. $\l$ gives the well-known equation
\begin{equation}
\Re\left(\omega_0(\l)\right)\ =\ \Pint_{\cal C} \D\mu\,
\rho(\mu)\,{1\over \l-\mu}\ =\  V'(\l)/2\qquad \forall\, \l\in{\cal C}
     \label{eq:spres}
\end{equation}
where $\omega_0$ is the large $N$ resolvent
\begin{equation}
      \omega_0(\l)\ =\ \lim_{N\to\infty}\langle{1\over N}
      \Tr\left[{1\over\l-M}\right]\rangle\ =\ \
      \int_{\cal C} \D\mu\,{\rho(\mu)\over\l-\mu}
     \label{eq:resolv}
\end{equation}
Finally let us recall that in the one-cut case ${\cal C}=[a,b]$,
if the potential $V$ is a
polynomial of degree $P$, $\omega$ is of the form
\begin{equation}
      \omega_0(\l)\ = {V'(\l)\over 2}\,-\,{M(\l)\sqrt{\sigma(\l)}\over 2}
      \with\sigma(\l)\ =\ (\l-a)(\l-b)
     \label{eq:omeonecut}
\end{equation}
where $M(\l)$ is a polynomial with degree $P-2$.
$a$, $b$ and $M$ are entirely determined by the constraint that
\begin{equation}
      \omega_0(\l)\ = \l^{-1}\,+\,{\cal O}(\l^{-2})\for\l\to\infty
     \label{eq:omeginfty}
\end{equation}
The e.v. density is given by the discontinuity of $\omega$
\begin{equation}
      \rho(\l)\ =\
      {\ii\over 2\pi}\left[\omega(\l+\ii 0_+)-\omega(\l-\ii 0_+)\right]
      \ =\ {M(\l)\sqrt{|\sigma(\l)|}\over 2\pi}
     \label{eq:rhoexp}
\end{equation}

\subsection{The 2-cut case}
\label{subsubsec.2.2}
\subsubsection{Mean field:}
\label{subsubsec.2.2.1}
If the potential $V$ is real but has more than one minimum, the large
$N$ limit may be described by an e.v. distribution on several
disconnected intervals.  For simplicity we shall first consider the
case where there are two intervals
\begin{equation}
      {\cal C}\ ={\cal C}_1\cup{\cal C}_2
      \quad ,\quad
      {\cal C}_1=[a,b]\quad,\quad{\cal C}_2=[c,d]\quad,\quad a<b<c<d
     \label{eq:2int}
\end{equation}
In this case, as we shall see,
\textbf{there is no topological large $N$ expansion},
even for the free energy $F$.
As shown in \cite{Jurk90,David91}, to describe the large $N$ limit, we
have to consider as an additional variable the "average" proportion of
eigenvalues $x_1=n_1/N$ and $x_2=n_2/N$ in each interval ${\cal C}_1$ and
${\cal C}_2$, and introduce the associated Lagrange multipliers
$\Gamma_1$ and $\Gamma_2$ for the constraints
\begin{equation}
      x_\alpha\,=\,\int_{{\cal C}_\alpha}\!\rho(\l)\,\D\l\
      \quad,\quad \alpha\,=\,1,2
     \label{eq:xconstr}
\end{equation}
The effective action \ref{eq:barS} now reads, with
\begin{equation}
     x=x_1
     \label{eq:xdef}
\end{equation}
\begin{equation}
      {\bar S}[\rho;x]\ =\ S[\rho]\,+\, \sum_{\alpha=1}^{2}\,\Gamma_\alpha\,
      \big(x_\alpha-\int_{{\cal C}_\alpha}\!\rho(\l)\,\D\l\big) \quad,\quad
      x_1+x_2=1
     \label{eq:S2cut}
\end{equation}
with $S[\rho]$ given by \ref{eq:Srho} as before.  The saddle point
equation w.r.t. $\rho(\l)$ gives as before the equation
\ref{eq:spequ}, which implies that the effective potential defined by
\ref{eq:Veff} is constant on each interval
\begin{equation}
     V_{\rm eff}(\l) = \Gamma_\alpha \qquad{\rm when}
     \qquad \l\in {\cal C}_\alpha
     \label{eq:veff2c}
\end{equation}
but the corresponding e.v. density $\rho_c(\l)$ and the effective action
${\bar S}_c$ still depend explicitly of the e.v. proportion $x$,
since we have
\begin{equation}
     {\bar S}_c[x]\ =\ {1\over 2}\left(\int_{\cal C}\rho_c(\l)\,V(\l)\,\D\l\,+
     \,\sum_\alpha \Gamma_\alpha x_\alpha\right)
     \label{eq:Sc2c}
\end{equation}
The saddle point equation w.r.t. $x$
implies the equality of the effective potentials for each interval
\begin{equation}
     {\partial{\bar S}\over\partial x}\,=\,\Gamma_1-\Gamma_2\ =\ 0
     \label{eq:sdlpteqx}
\end{equation}
This fixes the value of $x$, and it is known that with this last
equation the e.v. density $\rho_c$ is uniquely determined in the 2-cut case.
The large $N$ free energy is then given simply by
\begin{equation}
     F_0\ =\ {\bar S}[\rho_c;x_c]\ =\ {\bar S}_c[x_c]
     \label{eq:F02cut}
\end{equation}

For an explicit polynomial potential $V$ of degree $P$, and for fixed $x$,
the 2-cut mean field solution for the resolvent is
\begin{equation}
     \omega_0(\l,x)\,=\, {V'(\l)\over 2}\,-\,{M(\l)\sqrt{\sigma(\l)}\over 2}
     \with
     \sigma(\l)\,=\,(\l-a)(\l-b)(\l-c)(\l-d)
     \label{eq:omtwocut}
\end{equation}
and $M(\l)$ a polynomial with degree $P-3$.
The e.v. density $\rho(\l,x)$ is still given by the discontinuity of 
$\omega_0$.
The coefficients of $M$ and the 4 end-points $a,b,c,d$ are entirely
determined by the constraint that $\omega_0(\l)\simeq\l^{-1}$ when
$\l\to\infty$ and by the fact that $x$ must be given by
\begin{equation}
     x\ =\ \int_{a}^{b}\rho(\l,x)\,\D\l\,=\,
     {1\over 2\pi}\int_{a}^{b}|M(\l,x)|\sqrt{|\sigma(\l)|}\,\D\l
     \label{eq:xdefint1}
\end{equation}
Finally the equation \ref{eq:sdlpteqx} which fixes $x=x_{c}$ reads
\begin{equation}
     0\,=\,V_{\rm eff}(b)-V_{\rm eff}(c)
     \,=\,
     \int_{b}^{c}\D\l\,\left(2\omega_{0}(\l,x)-V'(\l)\right) \,=\,
     -\int_{b}^{c}\D\l\,M(\l,x)\sqrt{|\sigma(\l)|}
     \label{eq:equeffpot}
\end{equation}

\subsubsection{Discreteness of number of e.v.'s:}
\label{subsubsec.2.2.2}
This is sufficient
if one is interested in the leading term in the large $N$ limit
(planar approximation).  However, in order to understand the structure
of the subdominant terms of the large $N$ expansion, it turns out that
we cannot neglect the fact that the number of e.v. $n_\alpha=N
x_\alpha$ in each interval ${\cal C}_\alpha$ must be an integer.

Let us consider the simple case where the potential $V$ has two separate
minima $z_{1}$ and $z_{2}$.
What has to be done is first to fix the
number of eigenvalues $n_{1}=n$ (resp.  $n_{2}=N-n$) in the vicinity of
$z_{1}$ (resp.  $z_{2}$) in the partition function \ref{eq:evZ} by
writing
\begin{equation}
     \tilde Z[V;N]\ =\ \sum_{n=0}^N\,{N!\over n!(N-n)!}\,\tilde Z[V;n,N-n]
     \label{eq:ZZn}
\end{equation}
where
\begin{equation}
     \tilde Z[V;n,N-n]\ =\
     \int_{-\infty}^{E}\prod_{i\le n}^{}\D{\l_i}
     \int_{E}^{+\infty}\prod_{j>n}^{}\D{\l_j}
     \,\, \ee{-N \sum_k V(\l_k)}\,\, \prod_{k<l}
     (\l_k-\l_l)^2
     \label{eq:Zn}
\end{equation}
with $E$ a "frontier" $b<E<c$ between the two semi-classical cuts $[a,b]$
and $[c,d]$.
We now claim that each term of this discrete sum has a well defined large $N$
topological
expansion.
Indeed, we can rewrite
\ref{eq:Zn} as a matrix integral over two separate matrices: a
$n_{1}\times n_{1}$ matrix $M_{1}$ with the $n_{1}=n$ e.v. $<E$ and a
$n_{2}\times n_{2}$ matrix $M_{2}$ with the $n_{2}=N-n$ e.v. $>E$, as
\begin{equation}
     \tilde Z[V;n]\ =\ {1\over C_{n}C_{N-n}}\,
     \int \dmat{n_{1}}{M_{1}}\,\int \dmat{n_{2}}{M_{2}}\e^{-N\Tr\left(V(M_{1})
     \right)
     -N\Tr\left(V(M_{2})\right)
     +2\Tr\left(\ln\left(M_{1}\otimes\mathrm{Id}+\mathrm{Id}\otimes
     M_{2}\right)\right)}
     \label{eq:2matrint}
\end{equation}
This last matrix integral has a topological large $N$ expansion of the
form \ref{eq:topexp} in the `t~Hooft limit $N\to\infty$, $x=n/N$ fixed,
obtained by doing a classical perturbative expansion around the
smallest minimum $z_{1}$ of $V$ for $M_{1}$ and around the largest
minimum $z_{2}$ of $V$ for $M_{2}$, and by re-organizing the
perturbative expansion according to the topology of the Feynman
diagrams.  Taking into account carefully the measure factors $C_{n}$
and $C_{N-n}$, and using their large $N$ asymptotics
\begin{equation}
     C_{N}\ =\ {1\over N!}\,\left({2\pi\over N}\right)^{{N^2\over 2}}\,
     \e^{{3\over 4}N^2} \, (2\pi)^{-N}\,N^{1\over
     12}\,\mathrm{cst}\,(1+{\cal O}(N^{-1})) \hbox{ when } N\to\infty
     \label{eq:Cnasympt}
\end{equation}
(easily derived from Stirling formula),we obtain that
\begin{equation}
     Z[V;N]\ =\ \left({2\pi\over N}\right)^{{N^2\over 2}}\,N^{-{1\over 12}}
     \,\sum_{n=0}^{N}\e^{-F[V;N,x]}
     \label{eq:ZNnsum}
\end{equation}
where each $F[V;N,x]$ has a regular large $N$ asymptotic expansion of the form
\begin{equation}
      F[V;N,x]\ =
      \ \sum_{h=0}^\infty\,
      N^{2-2h}\,F_h[V,x] \where x\,=\,n/N
     \label{eq:Fnexp}
\end{equation}
with each $F_h[V,x]$ a regular function of $x=n/N$.
In particular, the leading large $N$ term is given (up to an additive
-- $V$ \textbf{and} $x$ independent -- constant) by the classical
effective action \ref{eq:F02cut}
\begin{equation}
     F_0\ =\ {\bar S}[\rho_c;x_c]\ =\ {\bar S}_c[x_c]
     \label{eq:F02cutbis}
\end{equation}
Finally, let us stress that although this decomposition depends on the
arbitrary parameter $E$, since $E$ is in the interval $]b,c[$ where
the density of eigenvalues is exponentially small with $N$, the integral
\ref{eq:Zn} depends on $E$ only through exponentially small terms of order
$\ee{-\mathrm{cst}\cdot N}$, which are ``non-perturbative" in the
topological expansion \eq{eq:Fnexp}.

\subsubsection{Beyond mean-field:}
\label{subsubsec.2.2.3}
We can now easily calculate the subleading terms of order
${\cal O}(N^{-2})$ for the full partition function.
In the large $N$ limit we can approximate the sum \ref{eq:Fnexp} by
\begin{equation}
     Z[V;N]\ \propto\ \sum_{n=0}^{N}\e^{-N^{2}F_{0}[V;x]-F_{1}[V;x]+\cdots}
     \label{eq:Z2csub}
\end{equation}
If $x_{c}$ denotes the saddle point of $F_{0}[x]$ given by
\ref{eq:sdlpteqx}, the sum is dominated by the $n$'s such that
\begin{equation}
     \vert n-N x_{c}\vert \ =\ {\cal O}(1)
     \label{eq:ndom}
\end{equation}
Thus we can still use a quadratic approximation for $F_{0}[x]$
\begin{equation}
     Z[V;N]\ \propto\
     \e^{-\left(N^{2}F_{0}[V;x_{c}]+F_{1}[V;x_{c}]\,+\,\cdots\right)}\
     \sum_{n}^{}\ee{-\,(n-N x_{c})^{2} F''_{0}[V;x_{c}]/2}
     \label{eq:Z2cquad}
\end{equation}
where $F''_{0}=\partial^{2}F_{0}/\partial x^{2}$ and where the $\cdots$
represent terms of order ${\cal O}(N^{-2})$.  The last sum over
$n$ gives simply an elliptic Jacobi theta function $\theta_{3}$
\begin{equation}
     \sum_{n}^{}\ee{-(n-N x_{c})^{2} F''_{0}[V;x_{c}]/2}
     \ =\
     \left(2\pi F''_{0}[V;x_{c}]\right)^{-1/2}\,\theta_{3}(N x_{c}\vert\tau)
     \label{eq:sum2theta}
\end{equation}
with modular parameter $\tau$ given by
\begin{equation}
     \tau\ =\ {2\ii\pi\over  F''_{0}[V;x_{c}]}
     \label{eq:deftau}
\end{equation}
and where the theta function is defined as
\begin{equation}
     \theta_{3}(z\vert\tau)\,=\,\theta_{3}(z)\,= \sum_{n\in{\mathbb
     Z}}q^{n^{2}}\ee{2\ii\pi n z} \with q=\ee{\ii\pi\tau}
     \label{eq:deftheta}
\end{equation}
It obeys the periodicity relations
\begin{equation}
    \theta_{3}(z+1)=\theta_{3}(z)
    \quad,\quad
    \theta_{3}(z+\tau)=\ee{-\ii\pi(2 z+\tau)}\theta_{3}(z)
\end{equation}
(For details on elliptic functions see e.g. Refs
\cite{Bateman,Abra,Elliptic}).
Eventually we have for
the free energy
\begin{eqnarray}
     \begin{array}{rcl}
     F[V;N]\ =\ &N^{2}&\,F_{0}[V,{x}_{c}]\\
     &\, -&\, \ln\left({\theta_{3}(N{x_c})}\right) \,+\,F_1(V;{x}_c) \,+\,
     {1\over 2}\,\ln\left({2\pi F''_{0}[V;x_{c}]}\right)\\
     &\,+&\,{\cal O}(N^{-2})
     \end{array}
     \label{eq:F2cfin}
\end{eqnarray}
$$ $$
where $F_1$ is the torus contribution in the topological expansion of
\ref{eq:2matrint}.
The next terms of this expansion can be calculated along the same line.

Let us stress that this is not a topological expansion, since the second term
$\ln\left({\theta_{3}(N{x_c})}\right)$, seemingly ${\cal
O}(1)$ and contributing at the torus order, is \textbf{not regular} in
$N$.  Indeed, it is periodic in $x_{c}$ with period $1/N$.  When
computing some observables or quantities of the matrix model, one must
take derivatives of $F$ w.r.t. some parameters of the potential $V$.
Since the saddle point $x_{c}$ depends implicitly on $V$, every
derivative will give a factor $N$, and this term may become of the
same order than the first term $N^{2}F_{0}[{x}_{c}]$ given by the planar limit.
Note that the last two terms depend on ${x}$ and not on $N{x}$, and
they will remain subdominant once we take derivatives of $F$.
\subsection{The modular parameter}
\label{subsec.2.4}
Finally, we can express simply the modular parameter $\tau$
defined by \eq{eq:deftau} in term of the end-points $a$, $b$, $c$, $d$
of the support of e.v.
For this purpose, we introduce the function
$\sigma$
\begin{equation}
     \sigma(\l)\ =\ (\l-a)(\l-b)(\l-c)(\l-d)
     \label{eq:sig2cut}
\end{equation}
and the function $u$
\begin{equation}
     u(\l) = {1\over 2K }\int_d^\l {\D{z}\over\ssq{z}}
     \label{eq:udef}
\end{equation}
where $K$ is
\begin{equation}
     K\, =\, \int_b^c {\D{z}\over\sqrt{|\sigma(z)|}}\,=\,
     {2\over\sqrt{(c-a)(d-b)}}\,K[m]
     \with m\,=\,{(d-a)(c-b)\over (d-b)(c-a)}
     \label{eq:Kexpl}
\end{equation}
$K[m]$ is the complete elliptic integral of the first kind.  Similarly
we define
\begin{equation}
     K'\,=\,\int_a^b {\D{z}\over\sqrt{|\sigma(z)|}}\,=\,
     {2\over\sqrt{(c-a)(d-b)}}\,K[m']
     \with m'\,=\,1-m
     \label{eq:K'expl}
\end{equation}

We shall
show that the modular parameter $\tau$ of \eq{eq:deftau} coincides
with the standard modular parameter of the torus associated to the
mapping $u$, i.e. of the elliptic curve $y^{2}=\sigma(z)$.
Indeed, $\tau$ is simply given by
\begin{equation}
     \tau\ =\ \ii{K'\over K}\ =\ \ii{K[1-m]\over K[m]}\
     \label{eq:tauK}
\end{equation}
So we have
\begin{equation}
     u(d)=0 \hspace{19pt}, \hspace{19pt}
  u(a)={1\over2} \virg u(b)={1+\tau\over 2}
     \hspace{19pt}, \hspace{19pt} u(c)={\tau\over 2}
\hspace{19pt}, \hspace{19pt} u(\infty)=u_\infty
     \label{eq:uper}
\end{equation}
and $u$ maps the upper half $\l$-plane onto the half-periods rectangle
$(1/2,\tau/2)$ and the double-sheeted complex $\l$-plane onto the
period rectangle $(1,\tau)$.
\begin{figure}[tbp]
\vspace{0.5cm}
\centering
{\epsfxsize=15cm\epsfbox{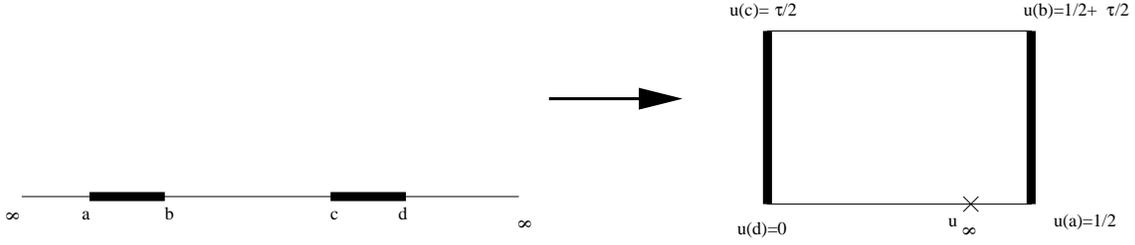}}
\vspace{0.5cm}
\caption{the upper half-plane is mapped onto a rectangle ($1/2,\tau/2$)}
\label{fig:1}
\end{figure}

To show \eq{eq:tauK}, we use the fact that in the two-cut case, if we
fix $x$ (the e.v. ratio in the first cut) the semiclassical e.v.
density (extrema of the effective action ${\bar S}$ ) is now a
function $\rho(\l,x)$ of $\l$ and $x$, and the end-points $a$, $b$,
$c$, $d$ depend on $x$.  Therefore the large $N$ resolvent
$\omega_{0}$ is of the form
\begin{equation}
     \omega_0(\l,x)\ = {V'(\l)\over 2}\,-\,{M(\l,x)\sqrt{\sigma(\l)}\over 2}
     \label{eq:omeg2c}
\end{equation}
with $M(\l,x)$ a polynomial with degree $P-3$ in $\l$
($P$ being the degree of $V$), entirely fixed by the constraints
\ref{eq:omeginfty} and \ref{eq:xdefint1}.
     Therefore the partial derivative of
$\omega_0(\l,x)$ w.r.t. $x$ is necessarily of the form
\begin{equation}
     {\partial\omega_0(\l,x)\over\partial x}\,=\,
     {C\over\sqrt{\sigma(\l)}}
     \label{eq:domegdx}
\end{equation}
with $C=C(\l,x)$ a priori a polynomial in $\l$.  Since \ref{eq:omeginfty}
still holds independently of $x$ we must have
\begin{equation}
      {\partial\omega_0(\l,x)\over\partial x}\,=
      \,{\cal O}(\l^{-2})\for\l\to\infty
     \label{eq:domegdxinfty}
\end{equation}
which implies that $C(\l,x)$ is of degree $0$ in $\l$, i.e. is a constant
(depending only on $x$)
\begin{equation}
     C\,=\,C(x)
     \label{eq:Cindep}
\end{equation}
This constant can be easily determined by using that
\begin{equation}
     x\ =\ \int_{a}^{b}\rho(\l,x)\,\D\l\,=\,
     \int_{\cal C'}{\D\l\over 2\ii\pi}\omega_0(\l,x)
     \label{eq:xdefint}
\end{equation}
with ${\cal C'}$ a clockwise contour encircling the interval $[a,b]$.
Therefore we have
\begin{equation}
     {\partial x\over\partial x}\,=\,1\,=\,
     \int_{\cal C'}{\D\l\over 2\ii\pi}{C\over\sqrt{\sigma(\l)}}
     \,=\, -{CK'\over\pi}
     \quad\Rightarrow\quad C\,=\,-{\pi\over K'}
     \label{eq:CKprime}
\end{equation}
with $K'$ the half-period defined in \eq{eq:K'expl}.
Now we use \eq{eq:sdlpteqx}, \eq{eq:F02cut} and the definition of the
effective potential $V_{\rm eff}$ of \eq{eq:Veff} to write the
derivative of the free energy w.r.t. $x$ as
\begin{equation}
     {\partial F_{0}\over\partial x}\,=\,V_{\rm eff}(b)-V_{\rm eff}(c)
     \,=\,
     \int_{b}^{c}\D\l\,\left(2\omega_{0}(\l)-V'(\l)\right)
     \label{eq:dFdx}
\end{equation}
Now we take the derivative w.r.t. $x$ of this equation and obtain
\begin{equation}
     F_{0}''\,=\, {\partial^{2} F_{0}\over\partial x^{2}}\,=\,
     2\,\int_{b}^{c}\D\l\,{\partial \omega_{0}(\l)\over\partial x} \,=\,
     2\,\int_{b}^{c}\D\l\,{C\over\sqrt{\sigma(\l)}} \,=\, -\,{2CK}
     \,=\,{2\pi K\over K'}
     \label{eq:d2Fdx2}
\end{equation}
Using \eq{eq:deftau} we thus obtain the result \ref{eq:tauK}.


\newsection{2-points correlation function}
\label{sec.3}
\subsection{The basic formula}
\label{subsec.3.1}
As a first application we compute the large $N$ smoothed connected two-point
correlation function (first obtained by
\cite{kanz98,BrezinDeo,deift}), defined as
\begin{equation}
     \omega_{}^{c}(\lambda,\mu)\ =\ 
\langle\Tr\left[{1\over\lambda-M}\right]\Tr\left[{1\over\mu-M}\right]\rangle
     \,-\,
     \langle\Tr\left[{1\over\lambda-M}\right]\rangle
     \langle\Tr\left[{1\over\mu-M}\right]\rangle
     \label{eq:2ptFunct}
\end{equation}
Adding  source terms to the potential of the form
\begin{equation}
     V_{\epsilon_{\lambda}}\,=\,
     V(z)\,-\,\epsilon_{\lambda}{1\over\lambda-z}
     \quad,\quad V_{\epsilon_{\lambda},\epsilon_{\mu}}(z)\,=\,
     V(z)\,-\,\epsilon_{\lambda}{1\over\lambda-z}
     \,-\,\epsilon_{\mu}{1\over\mu-z}
     \label{eq:Vpluseps}
\end{equation}
we have
\begin{equation}
     \omega_{}^{c}(\lambda,\mu)\ =\
     -\,{1\over N^2}\,
     \left.{\partial\over\partial\epsilon_{\lambda}}
     {\partial\over\partial\epsilon_{\mu}}
     \,F[V_{\epsilon_{\lambda},\epsilon_{\mu}},N]
     \right|_{\epsilon_{\lambda}=\epsilon_{\mu}=0}
     \label{eq:2ptdeps}
\end{equation}
and for the resolvent (one-point function)
\begin{equation}
     \omega_{}^{}(\lambda)\ =\
     {1\over N}\,
     \langle\Tr\left[{1\over\lambda-M}\right]\rangle
     \,=\,-\,{1\over N^2}\,
     \left.{\partial\over\partial\epsilon_{\lambda}}
     \,F[V_{\epsilon_{\lambda}},N]
     \right|_{\epsilon_{\lambda}=0}
     \label{eq:1ptdeps}
\end{equation}
If the $\epsilon_{\lambda}$'s are small and the $\lambda$'s not too 
close to the
cuts, the mean-field solution is still a two-cut e.v. distribution,
with $x_{c}=x_{c}(\epsilon)$ an explicit function of the $\lambda$'s.
So from \eq{eq:F2cfin} for the free energy we have for the two-point function
\begin{equation}
     \omega_{}^{c}(\lambda,\mu)\ =\
     \left[
     -\,
     {\partial\over\partial\epsilon_{\lambda}}
     {\partial\over\partial\epsilon_{\mu}}
     \,F_{0}[V_{\epsilon_{\lambda},\epsilon_{\mu}}]
     \,+\,
     {\partial x_{c}\over\partial\epsilon_{\lambda}}
     \,
     {\partial
     x_{c}\over\partial\epsilon_{\mu}}
     \,
     \left[\ln\left({\theta_{3}(N{x_c})}\right)\right]''
     \right]_{\epsilon_{\lambda}=\epsilon_{\mu}=0} \,+\ {\cal O}(N^{-1})
     \label{eq:om2mf+thet}
\end{equation}
The first term in the r.h.s. of \ref{eq:om2mf+thet} is the mean-field
contribution already calculated in \cite{Ak96,AmAk}, the second term
involving a second derivative of an elliptic function, characterizes
the multi-cut solution.
\subsection{The mean-field contribution}
\label{subsec.3.2}
For completeness let us first rederive the mean field contribution of
\cite{Ak96,AmAk}.
Taking the derivative with respect to $\epsilon_{\lambda}$ we obtain the
mean-field resolvent for the potential $V_{\epsilon_{\mu}}$
\begin{equation}
     -\,\left.
     {\partial\over\partial\epsilon_{\lambda}}
     \,F_{0}[V_{\epsilon_{\lambda},\epsilon_{\mu}}]
     \right|_{\epsilon_{\lambda}=0}
     \,=\,
     \omega_{0}(\lambda;V_{\epsilon_{\mu}})
     \label{eq:mfres2cts}
\end{equation}
which must be of the form
\begin{equation}
     \omega_{0}(z;V_{\epsilon_{\mu}})
     \,=\,
     {1\over 2}\,\left[ V'(z) \,-\,{\epsilon_{\mu}\over (z-\mu)^{2}} \,+\,
     {M(z)\ssq{z}\over (z-\mu)^{2}} \right]
     \label{eq:omegepsmu}
\end{equation}
with $M(z)$ a polynomial of degree $P-1$ (here both the coefficients
of $M$ and of $\sigma$ depend on $\mu$ and $\epsilon_{\mu})$.
In addition to the $P-2$ constraints (coming from \ref{eq:omeginfty}
and \ref{eq:equeffpot}) $\omega_{0}(z;V_{\epsilon_{\mu}})$ must be
regular at $z=\mu$.  This determines entirely $M$. Taking the
derivative w.r.t. $\epsilon_{\mu}$ and using the symmetry 
$\l\leftrightarrow\mu$
we get
\begin{equation}
     \omega_{0}^{c}(\l,\mu) \,=\,-\,
     {\partial\over\partial\epsilon_{\lambda}}
     {\partial\over\partial\epsilon_{\mu}} \,F_{0}
     \,=\,
     -\,{1\over 2}\,{1\over (\l-\mu)^{2}}\,\left[
     1\,+\, {Q(\l,\mu)\over \ssq{\l}\ssq{\mu}} \right]
     \label{eq:om0explQ}
\end{equation}
with $Q(\l,\mu)$ a symmetric polynomial in $\l$ and in
$\mu$.
The constraints on $\omega_{0}^{c}$ are:
(i)  $\omega_{0}^{c}={\cal O}(\l^{-2})$ as $\l\to\infty$ which implies
that $Q$ is of degree at most 2;
(ii) $\omega_{0}^{c}$ is regular at $\l=\mu$ which implies that
$Q(\l,\mu)=-\sigma((\l+\mu)/2)+{\cal O}((\l-\mu)^{2})$;
(iii) finally the equality of the effective potential on the two cuts
implies that
\begin{equation}
	\int_{b}^{c}\omega_{0}^{c}(\l,\mu)\,\D\mu\,=\,0
	\label{eq:intom2c00}
\end{equation}
Conditions (i) and (ii) fix uniquely $Q$
\begin{equation}
	Q(\l,\mu)\,=\,-\,{1\over 2}\left[
	\begin{array}{c}
		\ (\l-a)(\mu-b)(\mu-c)(\l-d) \\
		+ (\mu-a)(\l-b)(\l-c)(\mu-d)
	\end{array}
	\right]
	\,+\,S\,(\l-\mu)^{2}
	\label{eq:Qexpl}
\end{equation}
up to a constant $S$ fixed by condition (iii)%
, which is found to be
\begin{equation}
     S\,=\,-\,{1\over 2}(c-a)(d-b){E[m]\over K[m]}
     \with m\,=\,{(c-b)(d-a)\over (c-a)(d-b)}
     \label{eq:SexplEK}
\end{equation}
and where $K[m]$ and $E[m]$ are the standard elliptic integrals of the
first and second kind.

\subsection{The non-regular contribution}
\label{subsec.3.3}
In order to compute the second contribution to \ref{eq:om2mf+thet}, we
simply need ${\partial x_{c}\over\partial\epsilon_{\l}}$.
Since $x_{c}$ is fixed by the constraint
${\partial F_{0}\over\partial x}=0$ we can write
\begin{equation}
     {\partial x_{c}\over\partial\epsilon_{\l}}\,=\, -\,
     {{\partial^{2}F_{0}\over\partial x\partial\epsilon_{\l}}\Bigg/
     {\partial^{2}F_{0}\over\partial x^{2}}}
     \label{eq:dcde}
\end{equation}
Using the results of subsection~\ref{subsec.2.4} we have
\begin{equation}
     {\partial^{2}F_{0}\over\partial x^{2}}
     \,=\,
     {2\pi K\over K'}
     \and
     {\partial^{2}F_{0}\over\partial x\partial\epsilon_{\l}} \,=\,
     -\,{\partial\hphantom{x}\over\partial x}\omega_{0}(\l,x) \,=\,
     {\pi\over K'}\,{1\over\sqrt{\sigma(\l)}}
     \label{eq:dfdxxdxe}
\end{equation}
So we have eventually
\begin{equation}
     {\partial x_{c}\over\partial\epsilon_{\l}}\,=\,
     -\,{1\over 2 K\sqrt{\sigma(\l)}}
     \label{eq:dcdeexpl}
\end{equation}
  and the second non-regular term is
\begin{equation}
     {\partial x_{c}\over\partial\epsilon_{\lambda}}
     \,
     {\partial
     x_{c}\over\partial\epsilon_{\mu}}
     \,
     \left[\ln\left({\theta_{3}(N{x_c}|\tau)}\right)\right]'' \,=\, {1\over 2
     K\sqrt{\sigma(\l)}} \, {1\over 2 K\sqrt{\sigma(\mu)}} \,
     \left[\ln\left({\theta_{3}(N{x_c}|\tau)}\right)\right]''
     \label{eq:sectermexpl}
\end{equation}
with $K$ defined by \eq{eq:Kexpl}.
Using standard relations on elliptic functions, this can be rewritten as
\begin{equation}
     {(c-a)(d-b)\over 4 \ssq{\l}\ssq{\mu}}\left[-{E[m]\over K[m]}+
     \dn^{2}(Nx_{c}+{\scriptstyle {1\over 2}})\right]
     \label{eq:sectermexpl2}
\end{equation}
with
\begin{equation}
     \dn(u)\,=\,\dn(2K[m]u|m)
     \label{eq:pqnorm}
\end{equation}
where $\dn(u|m)$ is the Jacobi elliptic function $\dn$.
Its periods are $2K[m]$ and $4\ii K[m']$, and $\dn^{2}(z)$ has periods
$1$ and $\tau$.

\subsection{The final result}
\label{subsec.3.4}
Combining \ref{eq:om0explQ}, \ref{eq:Qexpl},
\ref{eq:SexplEK} and \ref{eq:sectermexpl2} we obtain the final result
for the 2-point correlation function

\begin{eqnarray}
     \omega_{}^{c}(\l,\mu) \, & = &
     \,-\,{1\over 4(\l-\mu)^{2}}\,
     \left[\left(1-\sqrt{{(\l-a)(\l-b)(\mu-c)(\mu-d)\over
     (\mu-a)(\mu-b)(\l-c)(\l-d)}}\right)+
     \left(\l\leftrightarrow\mu\right)\right]\nonumber\\
          & & -\,{(c-a)(d-b)\over 4 \ssq{\l}\ssq{\mu}}\
          \sn^{2}(Nx_{c}+{\scriptstyle {1\over 2}})
      \label{eq:final2pt}
  \end{eqnarray}
We have used the relation $\dn^{2}(u)=1-m\,\sn^{2}(u)$, where similarly to
\ref{eq:pqnorm} we note
\begin{equation}
     \mathrm{sn}(u)=\mathrm{sn}(2K[m]u|m)
     \label{eq:snnorm}
\end{equation}
Surprisingly, the ratio $E[m]/K[m]$ characteristic of the mean-field
solution of \cite{Ak96,AmAk} has disappeared.

The smoothed 2-point connected density correlator $\rho^{c}(\l,\mu)$, 
defined as
\begin{equation}
     \rho^{c}(\l,\mu)\,=\,
     \langle\Tr\left[\delta({\lambda-M})\right]
     \Tr\left[\delta({\mu-M})\right]\rangle
     \,-\, \langle\Tr\left[\delta({\lambda-M})\right]\rangle
     \langle\Tr\left[{\delta(\mu-M})\right]\rangle
     \label{eq:2ptdens}
\end{equation}
can be obtained easily from the discontinuity of $ \omega_{}^{c}(\l,\mu)$.
One obtains in the large $N$ limit, if $\l$ and $\mu$ are on the 
support of e.v.
\begin{eqnarray}
     \rho^{c}(\l,\mu)\,&=\,&
     \,-\,{1\over 4\pi^{2}}
     \left[{1\over (\l-\mu)^{2}}\,
     \left(
         \sqrt{
	   \left|
                {(\l-a)(\l-b)(\mu-c)(\mu-d)\over
	       (\mu-a)(\mu-b)(\l-c)(\l-d)
	       }
	    \right|
	}
         \,+ \, \l\leftrightarrow\mu
     \right)
     \right.\nonumber\\
     & &
     \hphantom{\,-\,{1\over 4\pi^{2}}
     [}+\,\varepsilon_{\l}\varepsilon_{\mu}\, \left.{(c-a)(d-b)\over
     \sqrt{|\sigma(\l)|}\sqrt{|\sigma(\mu)|}}\
     \sn^{2}(Nx_{c}+{\scriptstyle {1\over 2}})
     \right]
     \label{eq:final2dens}
\end{eqnarray}
\begin{equation}
     \varepsilon_{\l}\,=\,1\quad\rm{if}\quad\l\in[a,b]\quad,\quad
     -1\quad\rm{if}\quad\l\in[c,d]\quad,\quad
     \label{eq:varepsdef}
\end{equation}
and zero otherwise.

The new non-regular term
$\sn^{2}(Nx_{c}+{\scriptstyle {1\over 2}})$
is an even periodic function of $Nx_{c}$ with period $1$
which varies between $0$ and $1$.  Therefore, as $N$ varies, depending on
the rationality or the irrationality of $x_{c}$, the two-point
function will be varying with $N$ in a periodic or quasiperiodic way.

\subsection{The symmetric case}
\label{subsec.3.5}
It is now very easy to recover the results of
\cite{kanz98,BrezinDeo} for a symmetric potential.  Indeed, if the 
potential $V$ is
symmetric, the two cuts are also symmetric
\begin{equation}
     a\,=\,-d\quad,\quad b\,=\,-c
     \label{eq:abcdsym}
\end{equation}
and we have automatically
\begin{equation}
     x_{c}\ =\ {1\over 2}
\label{eq:xhalf}
\end{equation}
so that
\begin{equation}
    \sn^{2}(Nx_{c}+{\scriptstyle {1\over 2}})
    \,=\,
    \left\{
    \begin{array}{lcc}
       \sn^{2}({\scriptstyle {1\over 2}})=1 \hbox{ if $N$ is even } &&\\
       \sn^{2}(0)=0 \hbox{ if $N$ is odd} &&
    \end{array}
%
     \right.
     \label{eq:snsym}
\end{equation}
\eq{eq:final2pt} and \eq{eq:varepsdef} become
  \begin{equation}
      \omega_{}^{c}(\l,\mu) \,  =
     \,-\,{1\over 2(\l-\mu)^{2}}\,
     \left[1-{(a^{2}-\l\mu)(b^{2}-\l\mu)\over\ssq{\l}\ssq{\mu}}
     \right]
     -\,{(-1)^{N}\over 2}\,{ab\over \ssq{\l}\ssq{\mu}}
       \label{eq:sym2pt}
  \end{equation}´
  \begin{equation}
      \rho_{}^{c}(\l,\mu) \, = \,
      {1\over 2\pi^{2}}
      \,{\varepsilon_{\l}\varepsilon_{\mu}\over
      \sqrt{|\sigma(\l)|}\sqrt{|\sigma(\mu)|}}
      \,\left( {(a^{2}-\l\mu)(b^{2}-\l\mu)\over(\l-\mu)^{2}}\,
      -\,(-1)^{N}\,ab\right)
       \label{eq:sym2dens}
  \end{equation}
with $\sigma(\l)=(\l^{2}-a^{2})(\l^{2}-b^{2})$.

\subsection{The two-point function as an elliptic function}
\label{subsec.3.6}

It is interesting to consider the two-point correlator in terms of the
elliptic coordinates defined by \eq{eq:udef}
\begin{equation}
     u=u(\l)\quad,\quad v=u(\mu)
     \label{eq:uv2lm+}
\end{equation}
Let us thus consider
\begin{equation}
     {\bar\omega}_{}^{c}(u,v)\,=\, {\partial \l\over\partial u}\,
     {\partial\mu\over\partial v}\,\omega_{}^{c}(\l,\mu)
     \,=\,
     2K\sqrt{\sigma(\l)}\,2K\sqrt{\sigma(\mu)}\,\omega_{}^{c}(\l,\mu)
     \label{eq:ombar2c+}
\end{equation}
It is easy to see (from the properties of $\omega_{}^{c}$) that
${\bar\omega}_{}^{c}(u,v)$ satisfies:
\begin{enumerate}
     \item ${\bar\omega}_{}^{c}(u,v)$ is a doubly periodic function of
     $u$ (and of $v$) with periods $1$ and $\tau$;

     \item   ${\bar\omega}_{}^{c}(u,v)$ is regular at $u$ and
     $v=u(a),\,u(b),\,u(c),\,u(d)$ and $u_{\infty}$;

     \item ${\bar\omega}_{}^{c}(u,v)$ is regular when $u=v$, but has a
     double pole at $u=-v$ (corresponding to the double pole of
     $\omega_{}^{c}(\l,\mu)$ when $\l=\mu$ but with $\l$ in the first
     sheet and $\mu$ in the second sheet), with residue $1$.
%

\end{enumerate}
This implies that ${\bar\omega}_{}^{c}(u,v)$ is a Weierstrass
elliptic function
\begin{equation}
     {\bar\omega}_{}^{c}(u,v)\,=\,\wp(u+v|\tau)+\mathrm{constant}
     \label{eq:weirthet+}
\end{equation}
where the constant depends on $Nx_{c}$ ($\wp$ has periods $1$ and $\tau$).
\iftrue Using classical identities between the Weirstrass $\wp$
function and the Jacobi elliptic functions, it can be easily
calculated.  \else Using the identities
\begin{eqnarray}
     \wp(u|\tau)+ 2\eta_{1}& = & \,-\left[\ln(\theta_{1}(u|\tau)\right]"
     \nonumber\\
     \eta_{1} & = & {\hbox{to be written}} \nonumber\\
     \theta_{1}(z|\tau) & = &-\ii\ee{\ii\pi(z+\tau/4)}
     \theta_{3}(z+{\scriptstyle{1+\tau\over 2}}|\tau)
     \label{eq:varioustheta}
\end{eqnarray}
with $\theta_{1}(z|\tau)$ the first Jacobi theta function, we can
compute the cst.
\fi
We find the remarkably simple result
\begin{equation}
     {\bar\omega}_{}^{c}(u,v)\,=\,\wp(u+v|\tau)\,-\,
     \wp(Nx_{c}+{\scriptstyle{\tau\over 2}}|\tau)
     \label{eq:2ptwpfin}
\end{equation} or equivalently
\begin{equation}
     {\bar\omega}_{}^{c}(u,v)\,=\,
     -\,\left[\ln\left({\theta_{1}(u+v|\tau)}\right)\right]''
     \,+\,\left[\ln\left({\theta_{3}(Nx_{c}|\tau)}\right)\right]''
     \label{eq:2ptth''}
\end{equation}

%
\newsection{The orthogonal polynomials}

Let us briefly recall some basic facts about the well-known method of
orthogonal polynomials \cite{polynomes}, which is a powerful tool for
studying the spectral properties of random matrices \cite{Mehta}.
Asymptotic expressions for the orthogonal polynomials have been
obtained recently \cite{deift} in the mathematical literature, by solving a
Rieman-Hilbert problem.
Here we will derive them from the free energy directly.

Consider the partition function \rf{eq:DM}:

\begin{equation}\label{eq:Zvp}
  \tilde{Z} = \int \D{\l_1}\dots\D{\l_N}  \,\, \ee{-N\sum_i V(\l_i)} \,\,
\prod_{i<j}
(\l_i-\l_j)^2
\end{equation}
The last term is a Vandermonde determinant \cite{Mehta}:
\begin{equation}\label{eq:vandermonde}
\prod_{i<j} (\l_i-\l_j) = \mathop{\det}_{i,j} \left( (\l_i)^{j-1}\right) =
\mathop{\det}_{i,j} \left( \calP_{j-1}(\l_i) \right)
\end{equation}
where the last equality is obtained by linearly mixing columns of the
determinant, and holds for arbitrary monic polynomials $\calP_n(\l)$ with
leading coefficient
$\calP_n(\l)=\l^n +\dots$.

The method of orthogonal polynomials consists in choosing a  family of
polynomials suitable for the computation of \rf{eq:Zvp},
namely, the family of polynomials
orthogonal with respect to the weight $\exp{-NV(\l)}$:
\begin{equation} \label{def:Polortho}
\int \D\l \,\,\calP_n(\l)\, \calP_m(\l) \, \ee{-NV(\l)} = h_n \delta_{nm}
\end{equation}

With this particular choice of polynomials, the integral \rf{eq:Zvp} is merely:
\begin{equation}\label{eq:Zorthpol}
  \tilde{Z}=  N!\,  \prod_{n=0}^{N-1} h_n
\end{equation}
and the joint probability density of all the eigenvalues takes the form of a
Slater determinant:
\begin{equation}\label{eq:probajoint}
R_N(\l_1,\dots,\l_N) = {1\over N!}\,\, \left( \mathop{\mathop{\det}_{0\leq
n<N}}_{1\leq
i\leq N} \, \left[\psi_{n-1}(\l_i)\right] \right)^2
\end{equation}
where the wave functions $\psi_n(\l) = {1\over \sqrt{h_n}} \calP_n(\l)
\ee{-{N\over 2}V(\l)} $ are orthonormal.

\subsection{The Kernel $K(\l,\mu)$}

The square of a determinant can be rewritten as the determinant of a product:
$$  \left( \mathop{\det}_{n,i} \left(\psi_{n-1}(\l_i)\right)\right)^2
  = \mathop{\det}_{1\leq
i,j\leq N}\left[ \sum_{n=0}^{N-1} \psi_n(\l_i)\psi_n(\l_j) \right] $$
we are thus led to introduce the kernel $K(\l,\mu)$ \cite{kernel}:
\begin{equation}\label{def:K}
  K(\l,\mu)= {1\over N} \sum_{n=0}^{N-1} \psi_n(\l)\psi_n(\mu)
\end{equation}
In terms of which the joint density of eigenvalues is now a determinant
\begin{equation}\label{eq:probjointker}
R_N(\l_1,\dots,\l_N) = {N^N\over N!} \det\left[ K(\l_i,\l_j)\right]
\end{equation}

The orthonormality properties of the polynomials imply the projection relations
\begin{equation}\label{eq:projker}
\int \D\l\, K(\l,\l)=1 \qquad {\rm and} \qquad \int \D\l\,
K(\mu,\l)K(\l,\nu)={1\over N}K(\mu,\nu)
\end{equation}
which make any partial integration of \rf{eq:probjointker} easy to perform
(theorem of Dyson \cite{Dyson}).

In particular, the integration over $N-1$ eigenvalues gives the density of
eigenvalues
$$ \rho(\l_1)= \int \D\l_2\dots\D\l_{N} R_N(\l_1,\dots,\l_N) \, = K(\l_1,\l_1)
$$
and the integration over $N-2$ eigenvalues gives the correlation function:
$$
\begin{array}{rl}
  R_2(\l_1,\l_2) & = \int \D\l_3\dots\D\l_{N} R_N(\l_1,\dots,\l_N) \\
  & = {N\over
N-1} \left( K(\l_1,\l_1)K(\l_2,\l_2)-K(\l_1,\l_2)K(\l_2,\l_1) \right)\\
\end{array} $$
In short:
\begin{equation}\label{eq:denscorrelker}
\rho(\l)=K(\l,\l) \qquad , \qquad \rho(\l,\mu) = \left(
K(\l,\l)K(\mu,\mu) - K(\l,\mu)^2 \right)
\end{equation}
In addition, the Darboux-Christoffel theorem \cite{Bateman,Eynard97},
asserts that
\begin{equation}\label{eq:Darboux}
  K(\l,\mu) = {1\over N h_{N-1}} { \calP_N(\l)\calP_{N-1}(\mu) -
\calP_N(\mu)\calP_{N-1}(\l) \over \l-\mu} \ee{-{N\over 2} (V(\l)+V(\mu))}
  \end{equation}
which means that we need to evaluate $\calP_n$ only for $n=N$ and $n=N-1$.

Thus, we shall now aim at finding asymptotic expressions for the orthogonal
polynomials $\calP_n(\l)$,
  and the kernel $K(\l,\mu)$ in the large $N$ limit, and $n$ close to $N$.
This has been done in the 1-cut case \cite{BrezinZee} and in the symmetric
2-cut case \cite{kanz98,deift,Deo}. Here we will generalize it to the 
non-symmetric case,
with the method used in \cite{Eynard97}.

\subsection{WKB approximation for the orthogonal polynomials $\calP_n(\l)$}

The orthogonal polynomials have the following integral representation (see
Appendix 1 of \cite{Eynard97} or \cite{polynomes}):
\begin{equation}\label{eq:Pnint}
  \calP_n(\l) = {\int \D{M}_{n\times n} \,\, \det{(\l-M)}\,\,\ee{-N\tr V(M)}
\over \int \D{M}_{n\times n} \,\, \ee{-N\tr V(M)} }
\end{equation}
where the integral is restricted to hermitian matrices of size $n\times n$.

Thus the orthogonal polynomial is given by the ratio of two matrix integrals of
the same type as the partition function \ref{eq:Z}:
\begin{equation}\label{eq:PnZ}
  \calP_n(\l) = {{Z[V+\delta V_1 +\delta V_2;n]}\over {Z[V+\delta
V_1;n]}  } = {\ee{-F[V+\delta V_1 +\delta V_2;n]}\over \ee{-F[V+\delta
V_1;n]}  }
\end{equation}
where
\begin{equation}\label{eq:defdV}
  \delta V_1 (z)= {N-n\over n} V(z) \and \delta V_2 = -{1\over n}\ln{(z-\l)}
\end{equation}

We have seen in the previous section (eq.\ref{eq:F2cfin}) that
\begin{equation}\label{eq:devFn}
  F[V;n] = n^2 F_0[V;x_c] -  \ln{\theta_3(nx_c[V])} + \dots
\end{equation}
We will use the fact that under a variation $\delta V$ of the potential, the
variation of $F_0$ is \cite{Eynard97,AmJuMa90}:
\begin{equation}\label{eq:dSdV}
  \delta F_0 = {1\over 2i\pi}\oint \om(z) \delta V(z) \,\D{z}
\end{equation}
where the anti-clockwise contour encloses the support of the density of
eigenvalues, and
$\om(z)$ is the resolvent (eq \ref{eq:1ptdeps} and \ref{eq:omtwocut}):
$$ \om(z)=  {1\over 2}\left( V'(z)
-M(z){\ssq{z}}\right) $$

It is convenient to introduce two sources $t_1$ and $t_2$ for the variations
$\delta V_1$ and $\delta V_2$ of the potential, and consider a generalized
potential
${\cal V}(z)$:
$$ {\cal V}(z) = V(z) + t_1 \delta V_1(z) + t_2 \delta V_2 (z) = V(z) 
+ t_1 V(z)
  + t_2 \ln{|\l-z|} $$
Since $t_1={N-n\over n} $ and $t_2=-{1\over n}$ are
both small of order $O(N^{-1})$, we will expand $F$ in Taylor's series:
$$ F[{\cal V}(z);n] = F[V(z);n] + t_1{\d_1 F} + t_2 {\d_2 F} + {t_1^2\over 2}
{\d_{11} F}  + {t_1 t_2} {\d_{12} F}  + {t_2^2\over 2} {\d_{22} F} +\dots$$
all the derivatives being taken at the point $t_1=t_2=0$.

This will give
\begin{equation}\label{eq:PndS}
  \calP_n(\l) \sim \ee{n{\d_2 F_0}} \ee{ (N-n){\d_{12} F_0}}  \ee{ -{1\over
2}{\d_{22} F_0} }\,\, {\theta_3(nx + (N-n)\d_1 x - \d_2 x) \over \theta_3(nx +
(N-n)\d_1 x)} \,\, (1  +O(N^{-1}))
\end{equation}

Now, let us compute the derivatives of $F_0$ and $x=x_c$
with respect to $t_1$ and $t_2$.
The method proceeds similarly to section \ref{subsec.3.1}.

\subsubsection{Derivatives of $F_0$ with respect to $t_1$ and $t_2$}

using \rf{eq:dSdV} with \rf{eq:defdV}:
$${\d F_0\over \d t_2} = {1 \over 2i\pi} \oint \om(z) \ln{(z-\l)} \D{z} $$
After integration by part, the pole in $(z-\l)$ picks a residue, and the result
is a primitive of $\om(\l)$:
\begin{equation}\label{eq:dSdttwo}
{\d F_0\over \d t_2} = \int_{\l_0}^\l \om(z) \D{z}
\end{equation}
The lower bound of integration $\l_0$ is to be chosen such that
$ \ee{n \d_2 F_0} \mathop{\sim}_{\l\to\infty} \l^n $.
i.e.
\begin{equation}\label{eq:lowerbound}
  \ln{\l_0} = \int_{\l_0}^\infty (\om(z)-{1\over z}) \D{z}
\end{equation}

In order to compute the second derivatives $\d_{12} F_0$ and $\d_{22} F_0$, we
will
need to differentiate $\om(z)$ with respect to $t_1$ and $t_2$.

\subsubsection{Derivatives of $\om(z)$ with respect to $t_1$ and $t_2$}

The resolvent $\om(z)$ computed for the potential ${\cal V}(z)$
takes the form:
\begin{equation}\label{eq:resolvgen}
  \om(z)={1\over 2} \left({\cal V}'(z) - M(z){\ssq{z}}\right)
\end{equation}
where $M(z)$ is analytic.
Notice that when ${\cal V}'(z)$ has a pole in $z=\l$, $M(z)$ may have a pole
too.
$\om(z)$ obeys a linear equation:
\begin{equation}\label{eq:colresolv}
  \om(z+i0)+\om(z-i0) = {\cal V}'(z)  \for z\in [a,b]\cup[c,d]
\end{equation}
Thus its derivatives obey linear equations as well:
\begin{equation}\label{eq:colderivtone}
  \d_1 \om(z+i0) + \d_1 \om(z-i0) = \delta V_1'(z) = V'(z)
\end{equation}
\begin{equation}\label{eq:colderivttwo}
  \d_2 \om(z+i0) + \d_2 \om(z-i0) = \delta V_2'(z) = {1\over z-\l}
\end{equation}

\noindent $\bullet$ $\d\om_1$:
The solution of \rf{eq:colderivtone} is:
\begin{equation}\label{eq:}
  \d_1 \om(z) = \om(z) - {f(z)\over {\ssq{z}}}
\end{equation}
where $f(z)$ is analytic in $z$.
The boundary conditions \ref{eq:omeginfty} imply that $f(z)\sim z$ when
$z\to\infty$ and $f$ has no pole, thus $f(z)$
is a polynomial of degree $1$:
\begin{equation}\label{eq:domdtone}
  {\d \om(z) \over \d t_1} = \om(z) - {z-z_0\over {\ssq{z}}}
\end{equation}
$z_0$ is determined as a function of $a,b,c,d$ by the derivative of
\ref{eq:equeffpot} with respect to $t_1$:
\begin{equation}\label{eq:zzero}
  \int_b^c \D{z} {z-z_0\over {\ssq{z}}} = 0
\end{equation}
It can be checked that in term of elliptic theta functions we have 
(see Appendix
A, or \cite{Elliptic}):
\begin{equation}\label{ephitheta}
  {z-z_0\over {\ssq{z}} } = {\D \over \D{z}} \ln{\theta_1(u(z)+u_\infty)\over
\theta_1(u(z)-u_\infty)}
\end{equation}
and thus:
\begin{equation}\label{eq:d1om}
  {\d \om(z) \over \d t_1} = \om(z) -  {\D \over \D{z}}
\ln{\theta_1(u(z)+u_\infty)\over
\theta_1(u(z)-u_\infty)}
\end{equation}

\bigskip

\noindent $\bullet$ $\d\om_2$:
Note that the $t_2$ source-term is the primitive of the $\epsilon_\l$
source-term of \ref{eq:Vpluseps}, and that
\begin{equation}\label{eq:domdttwo}
  {\D\over \D{z}} {\d \om(z) \over \d t_2} = {\d \om(z) \over \d\epsilon_\l}
=-{1\over n^2}{\d^2 F \over \d\epsilon_{z} \d\epsilon_\l} = \om_c(z,\l)
\end{equation}
so, $\d\om/\d t_2$ has already been computed in \ref{eq:om0explQ}.
The second derivative $\d_{22}F$ corresponds to $z=\l$.
\begin{equation}\label{eq:ddeuxS}
  \d_{22} F_0 = \ln{{\ssq{\l}}} + 2\ln{(\theta_1(u(\l)-u_\infty))}
\end{equation}

\subsubsection{Derivatives of $x$}

Recall that
\begin{equation}\label{eq:xdef1}
  x= \int_a^b \rho(\l) \D\l = {1\over 2i\pi} \int_a^b M(z){\ssq{z}} \D{z}
\end{equation}
using \ref{eq:d1om} we get:
\begin{equation}\label{eq:dxdtun}
  {\d x\over \d t_1} = x + {1\over 2i\pi} \int_a^b {z-z_0\over \ssq{z}} \D{z}
= x + 2 u_\infty
\end{equation}

Similarly, from \ref{eq:domdttwo} and \ref{eq:final2pt} or 
\ref{eq:2ptwpfin} (or
less tediously, taking the primitive of \ref{eq:dcdeexpl}), we get:
\begin{equation}\label{eq:dxdtdeux}
  {\d x\over \d t_2} = -u(\l) + u_\infty
\end{equation}

\subsection{Final result}

\subsubsection{Case $\l \in\kern -.75em/\, [a,b] \cup [c,d]$}

Eventually, inserting
\ref{eq:dSdttwo},\ref{eq:d1om},\ref{eq:ddeuxS},\ref{eq:dxdtun},
\ref{eq:dxdtdeux} into \ref{eq:PndS} we get:

\begin{equation}\label{eq:WKBnotinabcd}
  \calP_n(\l)
{\mathop{=}_{\l\notin [a,b]\cup[c,d]}}
  \sqrt{u'(\l)}\,\,
p_n(u(\l))
\,\, \ee{N\int^{\l}_{\l_0} \om}
\end{equation}

where
\begin{equation}\label{eq:Pn}
\encadremath{
  p_n(u) =   C_n \,
  {\theta_3(Nx + 2(N-n)u_\infty + u - u_\infty)\theta_1(2u_\infty)
\over \theta_3(Nx + 2 (N-n)u_\infty) \theta_1(u-u_\infty)}
    \left(  {\theta_1(u+u_\infty)\over \theta_1(u-u_\infty)}\right)^{n-N}
    }
\end{equation}
$C_n$ is a normalization such that $\calP_n \sim \l^n$ for $\l\to\infty$.
\begin{equation}\label{eq:defCn}
  C_n = \sqrt{2K} A^{n-N+1}
  \with A = -{1\over 2K}{\theta_1'(0)\over \theta_1(2u_\infty) } \and K=\int_c^b
{\D{z}\over\ssq{z}}
\end{equation}
$$ A=-{1\over 4}\,|d-a-c+b|\,\, {\theta_3(0)\over \theta_3(2 u_\infty)} $$

Note that \ref{eq:WKBnotinabcd} is unchanged under $u\to u+1$ and 
$u\to u+\tau$.
Indeed, a shift $u\to u+\tau$ amounts to a nontrivial circle around the cut
$[c,d]$.
Thus  $\int \om$ is shifted by $-2i\pi\int_c^d\rho = -2i\pi(1-x)$, and
$\ee{N\int\om}$ receives a phase $\ee{2i \pi N x}$.
In the same time, the $\theta$ functions receive phase factors:
$\theta(v+\tau) = \theta(v) \ee{-2i\pi (v +\tau/2)}$.
One can easily check that the total phase shift is $0$.

\subsubsection{Case $\l \in [a,b]\cup[c,d]$}

Expression \ref{eq:Pn} has been derived by a saddle point approximation of
\ref{eq:PnZ} when $\l$ does not belong to $[a,b]\cup[c,d]$.
When $\l$ lies on the cut $[a,b]\cup[c,d]$,
  \ref{eq:PnZ} actually has two saddle points, contributing to the same order.
They correspond to the two determinations of the square root $\pm \ssq{\l}$.
The asymptotic expression for the orthogonal polynomial is then given by a sum
of two terms:
\begin{equation}\label{eq:Pncut}
\calP_n(\l) {\mathop{=}_{\l\in [a,b]\cup[c,d]}} C \sqrt{u'}
\left[
    p_n(u) \, \ee{-iN\pi \zeta(\l) }
+i p_n(-u)\, \ee{ iN\pi \zeta(\l) }
\right]
  \,\,\ee{{N\over 2}V(\l)}
\end{equation}
where $\zeta(\l) = \int^{\l}_{d} \rho(z) \D{z}$
and
\begin{equation}\label{eq:defC}
  C=\ee{-{N\over2}\left(V(\l_0)+\int_{\l_0}^d M(z)\ssq{z}\D{z}\right)} = d^N
\ee{-{N\over 2}V(d)} \ee{-N\int_d^\infty (\om(z)-{1\over z}) \D{z}}
\end{equation}

\bigskip

To summarize:
When $\l\notin [a,b]\cup [c,d]$, the wave function $\psi_n(\l) = \calP_n(\l)
\ee{-NV/2}$ decays exponentially, and within the support $[a,b]\cup [c,d]$, it
oscillates at a frequency of order $N$.

\begin{figure}
\vspace{0.5cm}
\centerline{\epsfxsize=10cm\epsfbox{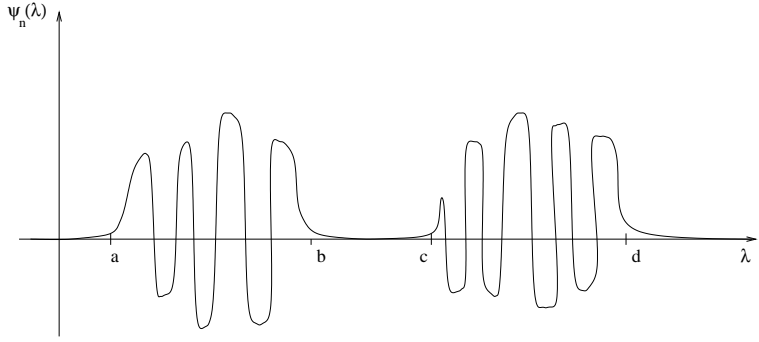}}
\vspace{0.5cm}
\caption{Typical behavior of the wave function}
\end{figure}

\subsubsection{Check of the orthogonality}

For completeness, let us check that the functions \rf{eq:WKBnotinabcd} are
indeed orthogonal (at leading
order in $N^{-1}$).
Let us compute the integral:
$$  \int_{-\infty}^\infty \D\l \calP_n(\l) \calP_m(\l) \ee{-NV(\l)} $$
The contributions of the integral along $]-\infty,a]\cup [b,c]\cup [d,\infty[$
are exponentially small and do not contribute at leading order.

Along $[a,b]\cup[c,d]$, we use expression \ref{eq:Pncut}, and get a sum of four
terms:
\begin{equation}\label{eq:checkpolortho}
C^2\, \int \D\l \,\, u'(\l) \left(
{\begin{array}{ll}
i  p_n(u) p_m(-u)
& +i  p_n(-u) p_m(u) \\
+  p_n(u) p_m(u) \ee{-2iN\pi\zeta(\l)}
& -  p_n(-u) p_m(-u) \ee{2iN\pi\zeta(\l)} \\
\end{array}}
\right)
\end{equation}
Since the two last terms have fast oscillations of frequency $N$, they are
suppressed as $O(1/N)$.

The leading contribution is thus given by the two first terms of
\ref{eq:checkpolortho},
which can be rewritten as integrals in the $u$ plane along the contour depicted
on fig.\ref{figdeform}.a:

\begin{figure}
\vspace{0.5cm}
\centerline{\epsfxsize=13cm\epsfbox{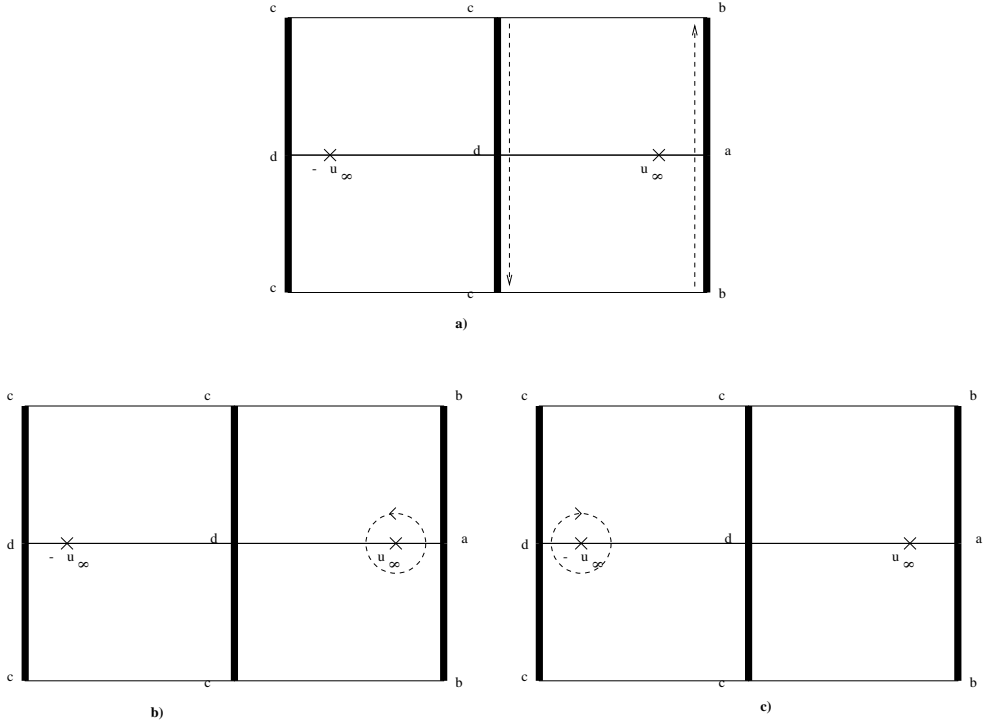}}
\vspace{0.5cm}
\caption{Deformation of the contour integral\label{figdeform}}
\end{figure}

\begin{eqnarray}
\label{eq:intpnpm}
{\begin{array}{ll}
\displaystyle \int \calP_n \calP_m \ee{-NV} =
  & \displaystyle i c_{nm} \int \D{u}
{\theta_3(x_n + u-u_\infty) \theta_3(x_m - u-u_\infty)\over
\theta_3(x_n)\theta_3(x_m)\theta_1(u-u_\infty)\theta_1(u+u_\infty)}
\left({\theta_1(u+u_\infty)\over\theta_1(u-u_\infty)}\right)^{n-m} \\
  & \displaystyle
+\,\, (u\to -u)\\
\end{array}}
\end{eqnarray}
where $x_n$ and $c_{nm}$ are short notations for:
\begin{equation}\label{eq:defxn}
x_n = N x + 2(N-n)u_\infty \and c_{nm} = C^2 C_n C_m \theta_1^2(2u_\infty)
\end{equation}

If $n>m$ we may deform the contour to a circle around the point $-u_\infty$
(fig.\ref{figdeform}.b), and the integral vanishes since there is no pole,
while if $m>n$ we deform the contour to a circle around $+u_\infty$
(fig.\ref{figdeform}.c).
Therefore, the integral vanishes for $n\neq m$.

When $n=m$, the integral picks a residue:

\begin{equation}\label{eq:ortho}
  \int \D\l \calP_n \calP_m \ee{-NV(\l)} = h_n \delta_{nm}
\end{equation}
with ($C$, $c_{nm}$, $A$, $x_n$ are defined in \ref{eq:defCn}, \ref{eq:defxn},
\ref{eq:defC}):
\begin{equation}\label{eq:hn}
  h_n = c_{nn} {4\pi  \over \theta_1(2u_\infty) \theta_1'(0)}
\,{\theta_3(x_{n+1})\over \theta_3(x_n)} =
  -{4\pi }\,
  C^2 \,\, {\theta_3(x_{n+1})\over\theta_3(x_n)} \,
A^{2(n-N+1/2)}
\end{equation}

\subsubsection{Recurrence equation}

It is well known that the orthogonal polynomials satisfy a recurrence equation
of the form \cite{Mehta,polynomes}:
\begin{equation}\label{eq:recurence}
  \l\calP_n(\l) = \calP_{n+1}(\l) + \beta_n \calP_n (\l) + \alpha_n
\calP_{n-1}(\l)
\end{equation}
Here, we find that (divide \ref{eq:recurence} by $\calP_n$, and match the poles
on both sides):
\begin{equation}\label{eq:eqalphan}
  \alpha_n = {h_n\over h_{n-1}} = {A^2}
{\theta_3(x_{n+1})\theta_3(x_{n-1})\over
\theta_3^2(x_n)}
\end{equation}
which can be rewritten more compactly as
\begin{equation}\label{eq:alphan}
  \alpha_n = {1\over 16}\left( ((d-a)-(c-b))^2 + 4 (d-a)(c-b)\,
\cn^2{(x_n+1/2)}\, \right)
\end{equation}

\bigskip

And by taking $u=0$ in \ref{eq:recurence} we get $\beta_n$:
\begin{equation}\label{eq:eqbetan}
  \beta_n-d = {A} \left[ {
\theta_3(x_{n+3/2})\theta_3(x_{n})\over \theta_3(x_{n+1})\theta_3(x_{n+1/2})}+{
\theta_3(x_{n+1})\theta_3(x_{n-1/2})\over
\theta_3(x_{n})\theta_3(x_{n+1/2})}\right]
\end{equation}
which can be rewritten more compactly as:
\begin{equation}\label{eq:betan}
  \beta_n = {a+d+(c-b)\over 2} - (c-b) {d-b\over c-b + {d-c\over
\cn^2{(x_n-u_\infty +1/2)} } }
\end{equation}

\bigskip

The sequences $\alpha_n$ and $\beta_n$ are thus quasi-periodic in $n$.
It is interesting to recall that the behavior of these coefficients has
been extensively studied (mainly by numerical methods) by several
authors in the early 90's \cite{chaosrm}.  The general conclusion was that in the
multi-cut case the general behavior of the recursion coefficients was
``chaotic'' in $n$ (and regular or quasi-periodic only in some special
cases). It is clear from our expressions that in the two-cut case
the behavior is \textbf{always} periodic or quasi-periodic and never chaotic
(in the mathematical sense).  This is in fact true even if the number
of cuts is larger than 2 (see appendix C).

In the symmetric case, $x={1\over 2}$ and $u_\infty={1\over 4}$, we have $x_n =
{n/2} \,\,{\rm mod}\, 1$, so that we recover
  $\beta_n=0$ and $\alpha_n = {1\over 4} (a-(-1)^n b)^2$.

In the general case, $\alpha_n$ and $\beta_n$ vary along a periodic curve,
between two extrema, given by:

$$  {(d-a - (c-b))^2 \over 16} \leq \alpha_n \leq {(d-a + (c-b))^2 \over 16} $$
$$  {d+a \over 2} - {c-b \over 2} \leq \beta_n \leq {d+a \over 2} + {c-b
\over 2}$$

Similarly to the one-cut case, one may relate $\alpha_n$ to square width of the
distribution of eigenvalues,
and $\beta_n$ to the center of the distribution.

\subsection{The kernel $K(\l,\mu)$}

We can now evaluate the kernel $K(\l,\mu)$ according to \rf{eq:Darboux}.
Let us note $u=u(\l)$ and $v=u(\mu)$ and we assume $\l,\mu \in [a,b]\cup[c,d]$:
\begin{equation}\label{eq:KPP}
K(\l,\mu) \sim {C^2 \sqrt{u' v'}  \over N h_{N-1}}
\sum_{\epsilon,\eta = \pm 1}
{ \sqrt{\epsilon\eta} p_N(\epsilon u) p_{N-1}(\eta v) \,\,\ee{-\epsilon Ni\pi
\zeta(\l)}\,\ee{ -\eta Ni\pi \zeta(\mu)}
- (u\to v) \over (\l-\mu)}
\end{equation}
  which can be rewritten as a sum of eight terms:
\begin{equation}
\label{eq:Ksum}
\begin{array}{cc}
\displaystyle
  K(\l,\mu) = &  \displaystyle {c_{N, N-1}\over h_{N-1} \theta_3(x_N)
  \theta_3(x_{N-1})}\,\,
{\sqrt{u'v'}\over N(\l-\mu)} \times\\
  & \displaystyle {\mathop{\sum}_{\epsilon,\eta,\kappa=\pm 1}}
  \kappa \sqrt{\epsilon\eta}
  {\theta_3(Nx+\epsilon u -\kappa u_\infty)\theta_3(Nx+\eta v +\kappa u_\infty)
  \over \theta_1(\epsilon u -\kappa u_\infty)\theta_1(\eta v +\kappa u_\infty)}
  \,\ee{-Ni\pi( \epsilon \zeta(\l)+\eta \zeta(\mu))} \\
\end{array}
\end{equation}

We will see below that not all the terms contribute to the same order.

\subsubsection{Regime $|\l-\mu| \sim O(1/N)$}

The eight terms of \rf{eq:Ksum} can be rewritten as
four combinations of the type:
$$ \sin{\left(  N\pi(\zeta(\l) \pm \zeta(\mu))\right)}
  {f(u,v) \mp f(v,u)\over N(\l-\mu)} \hspace{15pt}\makebox{and} \hspace{15pt}
  \cos{\left(  N\pi(\zeta(\l) \pm
\zeta(\mu)) \right)}
  {g(u,v) - g(v,u) \over N(\l-\mu)} $$
In the limit $|\l-\mu|$ small, i.e. $|u-v|$ small,
  the terms with a cosine will be proportional to derivatives of $g(u,v)$, and
there will be an overall ${1\over N}$ factor.
Similarly, the term with a sine and a $+$ sign will be proportional to a
derivative
of $f(u,v)$ and will be of order $1/N$.
Only the term proportional to $\sin{N\pi\int_\l^\mu \rho(z) \D{z}}$ can balance
the $1/N$ factor,
  and is dominant in the short range regime. After calculation we get:
\begin{equation}
  K(\l,\mu) {\mathop{\sim}_{|\l-\mu|\sim O(1/N)}} { \sin{N\pi\int_\l^\mu \rho(z)
\D{z}}\over N\pi(\l-\mu) }
\end{equation}

As expected we have
\begin{equation}
K(\l,\l)=\rho(\l)
\end{equation}
and we recover the universal short range correlation function:
\begin{equation}
\rho(\l,\mu) \sim \rho(\l)\rho(\mu) \left( 1-
\left({\sin{N\pi\rho(\l)(\l-\mu)}\over N\pi\rho(\l)(\l-\mu) }\right)^2\right)
\end{equation}

\subsubsection{Long range regime, smoothed oscillations}

When $|\l-\mu|\sim O(1)$, $K(\l,\mu)$ has high frequency oscillations, and
only a smoothed correlation function obtained by averaging the oscillations can
be observed.

Recall that the connected 2-point correlation function is related to $K^2$ by
\rf{eq:denscorrelker}:
\begin{equation}\label{eq:correlker}
   \rho_{2c}(\l,\mu) = - K(\l,\mu)^2
\end{equation}
with $K(\l,\mu)$ given by \ref{eq:Ksum}.

Smoothing out the oscillations amounts to kill all terms containing some
$\ee{iN\pi\zeta}$ in the square of eq \ref{eq:Ksum}, we thus have:

\begin{equation} \label{eq:Ksquaresmooth1}
\begin{array}{l}
\displaystyle
{\overline{K(\l,\mu)^2}} =  \displaystyle {-2 u'v' c_{N N-1}^2 \over h_{N-1}^2
\theta_3^2(x_N) \theta_3^2(x_{N-1}) N^2
(\l-\mu)^2}
\sum_{\epsilon,\eta,\kappa_1,\kappa_2=\pm 1}\,\,\kappa_1\kappa_2 \\
   \displaystyle {\theta_3(Nx+\epsilon u -\kappa_1 u_\infty)
   \theta_3(Nx-\epsilon u -\kappa_2
u_\infty)\theta_3(Nx+\eta v +\kappa_1 u_\infty)\theta_3(Nx-\eta v +\kappa_2
u_\infty)
\over
  \theta_1(\epsilon u- \kappa_1 u_\infty)\theta_1(-\epsilon u- 
\kappa_2 u_\infty)
  \theta_1(\eta v+ \kappa_1 u_\infty)\theta_1(-\eta v+ \kappa_2 u_\infty)
}\\
\end{array}
\end{equation}

Using that (see appendix A, and \cite{Elliptic,Abra,Bateman}):
\begin{equation}\label{eq:l-m}
  \l-\mu = -2A {\theta_1(u-v)\theta_1(u+v) \theta_1^2(2u_\infty) \over
\theta_1(u-u_\infty)\theta_1(u+u_\infty)\theta_1(v-u_\infty)\theta_1(v+u_\infty)
}
\end{equation}
we get ($\epsilon_{ij}=-\epsilon_{ji}=\pm 1$):

\begin{equation} \label{eq:Ksquaresmooth}
\begin{array}{l}
\displaystyle {\overline{K(\l,\mu)^2}} =
   \displaystyle {2 u'v' \theta_1'^2(0) \over N^2
4\pi^2\theta_3^4(Nx) \theta_1^2(2u_\infty)} \\ \displaystyle
\left({\theta_1(u-u_\infty)\theta_1(u+u_\infty)\theta_1(v-u_\infty)\theta_1
(v+u_
\infty) \over \theta_1(u-v)\theta_1(u+v)}\right)^2
  \displaystyle
\sum_{i,j,k,l=\pm 1}\,\,(\epsilon_{ij}\epsilon_{kl} +
\epsilon_{il}\epsilon_{kj}) \\
   \displaystyle {\theta_3(Nx+ u - i u_\infty)\theta_3(Nx- u -k u_\infty)
   \theta_3(Nx+ v - j
u_\infty)\theta_3(Nx- v - l u_\infty)
\over
  \theta_1(u- i u_\infty)\theta_1(- u- k u_\infty) \theta_1(v- j
u_\infty)\theta_1(- v- l u_\infty)
}\\
\end{array}
\end{equation}

We see that \ref{eq:Ksquaresmooth} has no pole when $u=\pm u_\infty$, it can
have (double) poles only when $u=\pm v$. Thus, \ref{eq:Ksquaresmooth} can be
rewritten in terms of Weirstrass functions of $u+v$ and $u-v$:

$$
{\overline{K(\l,\mu)^2}} = -{1\over 2N^2\pi^2}\, u'v'\,\left( C_1
\wp(u-v)+ C_2 \wp(u+v) -2 S \right)
$$
Taking $u=v$ and $u=-v$ in \ref{eq:Ksquaresmooth}, we find that the 
residues are
$C_1=C_2=1$, and taking a particular value of $u$ and $v$, we find the constant
$S$, equal to what we had in \ref{eq:2ptwpfin}:

\begin{equation}\label{eq:correl2ptsmooth}
{\overline{K(\l,\mu)^2}} = -{1\over 2N^2\pi^2}\, u'v'\,\left(
{\wp(u-v)+\wp(u+v)} -2 \wp(Nx+{\tau\over 2}) \right)
\end{equation}
and we recover the result \ref{eq:2ptwpfin} found in section \ref{subsec.3.6}.


\newsection{Conclusions}

In this article, we have solved the puzzle raised by
\cite{kanz98,BrezinDeo} and understood why the naive
mean-field method \cite{AmAk} and the orthogonal polynomial ansatz
\cite{kanz98,Deo} approach used in the symmetric case disagree.

We have proven here that this effect has nothing to do with a
$\mathbb{Z}_2$ symmetry breaking, as it was sometimes assumed
\cite{kanz98}, it is general as soon as the support of the density is
not connected.

The apparent paradox comes from the fact that when the support of eigenvalues
is not connected, the free energy admits no large $N$ expansion in powers of
$1/N^2$ (topological expansion \cite{BIPZ}).
This means that the free energy in the multi-cut case is not given by a
topological expansion, i.e. the sum of diagrams with a weight  $N^\chi$
($\chi$=Euler Characteristic of the diagram).

The explanation lies in the discreteness of the number of eigenvalues.
For instance in the symmetric 2-cut case, the classical approach assumes that
the minimum of the free energy is reached when one half ($x=1/2$) of the
eigenvalues are in each cut. Obviously, this minimum is never reached when
  the total number of eigenvalues is odd, and in general, the result depends
  on the fractional part of $Nx$.

At leading order in $N$ only, the free energy is correctly given by the
classical saddle point limit \cite{David91, AmAk}, but the first order in
$N$ is not sufficient to determine the 2-point (or higher) correlation
function.

Here we have computed explicitly the two-point connected correlation function.
It contains a universal part depending only on the number of cuts, which was
obtained by \cite{AmAk}, and contains in addition, a non universal term
quasiperiodic in $N$ \cite{chaosrm,deift}.

Let us stress that our calculation holds for any potential, not necessarily
symmetric, and it can also be generalized to a potential with complex
coefficients (appendix. B), and to an arbitrary number of cuts
(appendix. C).

We have also reobtained directly the asymptotic expressions for the 
orthogonal polynomials \cite{deift},
  which allows in principle through the Darboux-Christoffel theorem
(eq. \ref{eq:Darboux}) to compute any correlation function of any
number of eigenvalues in the short or long range domain (and one
can smooth it afterwards).

\bigskip

The orthogonal polynomial approach may in turn be used for other
random matrix ensembles, and it would be interesting to apply
our results to orthogonal or symplectic ensembles \cite{Mehta}.

\bigskip

The authors are thankful to K. Mallick for useful discussions, and to the
Eurogrid European Network HPRN-CT-1999-00161 for supporting part of the work.
They also thank E. Kanzieper, O. Lechtenfeld and G. Akemann for their
interest and for pointing some missing references.

\setcounter{section}{0}

%
%
\appendix{A few useful identities on elliptic functions}
\begin{figure}
\vspace{0.5cm}
\centerline{\epsfxsize=15cm\epsfbox{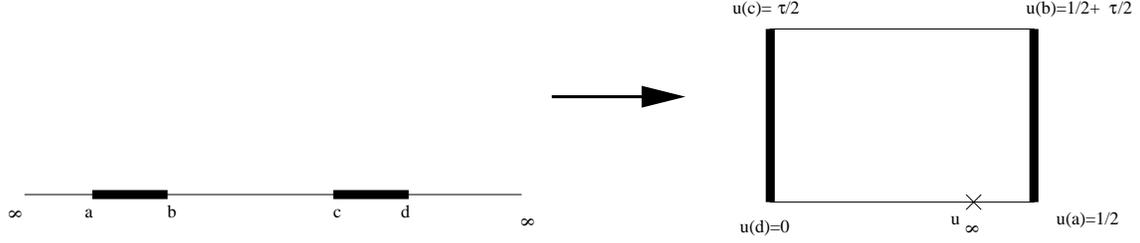}}
\vspace{0.5cm}
\caption{the upper half-plane is mapped onto a rectangle ($1/2,\tau/2$)}
\end{figure}

Here we collect a few useful identities on elliptic functions used
through the paper.  For details see \cite{Elliptic,Bateman,Abra}.
We start from
\begin{equation}
      \sigma(\lambda)\,=\,(\l-a)(\l-b)(\l-c)(\l-d)\qquad,\qquad a<b<c<d
      \label{eq:sig2}
\end{equation}
and the map from the complex plane to the torus
\begin{equation}
      u(\l) = {1\over 2K }\int_d^\l {\D{z}\over\ssq{z}}
      \label{eq:A2}
\end{equation}
where the half period $K$ is
\begin{equation}
      K\, =\, \int_b^c {\D{z}\over\sqrt{|\sigma(z)|}}\,=\,
      {2\over\sqrt{(c-a)(d-b)}}\,K[m] = {\pi\, \theta_3^2(0|\tau)\over
      \sqrt{(c-a)(d-b)}}
      \label{eq:K2A}
\end{equation}
$K[m]$ is the standard complete elliptic integral \cite{Elliptic,Bateman,Abra},
with the modulus $m$ equal to the biratio of the four points $a$, 
$b$, $c$, $d$:
\begin{equation}
      m\,=\,{(d-a)(c-b)\over (d-b)(c-a)}
      \label{eq:mbiratio}
\end{equation}
$m$ is related to the modular parameter $\tau$ of the torus by:
\begin{equation}
      m = \ee{i\pi\tau}\,{\theta_3^4({\tau\over2}|\tau)\over\theta_3^4(0|\tau)}
      \qquad {\rm and \,\, conversely} \qquad
      \tau\ =\ \ii{K[1-m]\over K[m]}\
      \label{eq:tauA}
\end{equation}
where we have used the Jacobi theta functions:
\begin{equation}
      \theta_{1}(z\vert\tau)\,=\,\theta_{1}(z)\,= -\ii\sum_{r\in{\mathbb
      Z+1/2}}(-1)^{r}\, q^{r^{2}}\,\ee{2\ii\pi r z} \with q=\ee{\ii\pi\tau}
      \label{eq:defthe1}
\end{equation}

\begin{equation}
  \and \theta_3(z\vert\tau)\,= \,\theta_3(z) \,=\, q^{1\over 4} \, 
\ee{i\pi z} \,
  \theta_1(z+{1\over 2}+{\tau\over 2} \vert \tau)
      \label{eq:defthe3}
\end{equation}

\medskip

With this mapping $u(\l)$ between the $\l$ complex plane and the periodic
rectangle of sides ($1,\tau$), we have:
\begin{equation}
      u(d)=0 \hspace{19pt} , \hspace{19pt} u(a)={1\over2} \hspace{19pt} ,
      \hspace{19pt} u(b)={1+\tau\over 2} \hspace{19pt} , \hspace{19pt} u(c)=
{\tau\over 2} \hspace{19pt} , \hspace{19pt} u(\infty)=u_\infty
      \label{eq:uqqch}
\end{equation}

The inverse mapping can be written in terms of theta functions:
\begin{equation}
      \l-d = -{\theta_1'(0)\over 2K} {\theta_1^2(u)\theta_1(2u_\infty )
      \over\theta_1(u+u_\infty )\theta_1(u-u_\infty )\theta_1^2(u_\infty )}
      \label{eq:Aone}
\end{equation}
\begin{equation}
      \ssq{\l} = {\theta_1'^2(0)\over 4K^2} {\theta_1(2u)\theta_1(2u_\infty )
\over\theta_1^2(u-u_\infty )\theta_1^2(u+u_\infty )}
      \label{eq:Atwo}
\end{equation}
and in terms of the usual trigonometric elliptic functions $\sn$, $\cn$,
$\dn$ \cite{Elliptic,Bateman,Abra} that we normalize to have periods
$1$ and $\tau$, i.e.
\begin{equation}
      \sn(u)\,=\,\sn(2K[m]u|m)\virg\dn(u)\,=\,\dn(2K[m]u|m)\virg\ldots
      \label{eq:elnorm}
\end{equation}
one has
\begin{equation}
      {\sn^2(u) \over \sn^2(u_\infty) } = {\l-d\over \l-c} \virg {\cn^2(u)
\over \cn^2(u_\infty )} = {\l-a\over \l-c} \virg {\dn^2(u) \over
\dn^2(u_\infty) } = {\l-b\over \l-c}
      \nonumber\label{eq:Athree}
\end{equation}
\begin{equation}
      \l-c = (d-c){1\over 1-{\sn^2{(u)}\over\sn^2({u_\infty) }}}
\virg \l-b = (d-b){\dn^2{(u)}\over 1-{\sn^2{(u)}\over\sn^2{(u_\infty )}}}
      \nonumber\label{eq:Afour}
\end{equation}
\begin{equation}
      \l-a = (d-a){\cn^2{(u)}\over 1-{\sn^2{(u)}\over\sn^2{(u_\infty) }}}
\virg \l-d = (d-c){d-a\over c-a}
{\sn^2{(u)}\over 1-{\sn^2{(u)}\over\sn^2{(u_\infty) }}}
      \nonumber\label{eq:Afive}
\end{equation}
\begin{equation}
      \ssq{\l} = (d-c)(d-a)\sqrt{d-b\over c-a}\,
{\sn{(u)}\,\cn{(u)}\,\dn{(u)}\over \left(1-{\sn^2{(u)}\over\sn^2{(u_\infty)
}}\right)^2}
      \nonumber\label{eq:Asix}
\end{equation}
and $u_\infty$ is related to $a,b,c,d$ by any of the following relations:
\begin{equation}
       \sn^2(u_\infty ) = {c-a\over d-a}
       \virg \cn^2(u_\infty ) = {d-c\over d-a}
       \virg \dn^2(u_\infty ) = {d-c\over d-b}
      \nonumber\label{eq:Aeight}
\end{equation}

\appendix{Complex potentials}
\label{app.B}
The case of complex potentials, that is to say of a polynomial potential
$V(\l)$ with complex coefficients, is interesting for some
applications of the matrix models to 2 dimensional gravity and when
studying their connections with integrable hierarchies.  In this case,
the mean field large $N$ solution is known to be given by a continuous
distribution of the eigenvalues along arcs in the complex plane
\cite{David93}.

In this appendix we show that our results are only slightly modified
in this case.

In the two-cut case, we can repeat the analysis of sect.\ref{sec.2}.  We fix
$x=n_{1}/N$ (the proportion of e.v. in the first cut). The resolvent is
still of the form \ref{eq:omtwocut}, with the polynomial $M$ and the end points
$a,b,c,d$ determined by the constraints \ref{eq:omeginfty} and
\ref{eq:xdefint1}, but they are no more real in general, as well as
the resulting mean-field free energy $F_{0}(x)$.

If we now repeat the calculation of sect.\ref{subsubsec.2.2.3} we cannot
use a saddle-point approximation for the sum over $n_{1}$ by expanding
$F_{0}(x)$ around the saddle point $x_{0}$ which is the true extremum
of $F_{0}$.
\begin{equation}
      {\partial F_{0}\over\partial x}(x_{0})\,=\,0
      \label{eq:x0col}
\end{equation}
Indeed this extremum is at a finite non-zero distance of the
real axis, i.e. $\Im(x_{0})={\cal O}(1)$, while the method of
sect.\ref{subsubsec.2.2.3} is valid only if $\Im(x_{0})={\cal O}(1/N)$.
However, since $N$ is integer, we can expand $F_{0}$ around any $x_{k}$,
provided that
\begin{equation}
      F'_{0}(x_{k})\,=\, 2\ii\pi\,{k\over N}\quad,\quad k\in{\mathbb Z}
      \label{eq:xkdef}
\end{equation}
since the dangerous oscillating term
$\ee{-N^{2}(x-x_{k})F'_{0}(x_{c})}$ is then a constant for $x=n/N$,
$n\in{\mathbb Z}$.

Therefore, as in \cite{David93} we have to consider the {\it real pseudo
saddle-point}  $x_{c}$ such that
\begin{equation}
     \Re(F'_{0}(x_{c}))\ =\ 0
     \with \Im(x_{c})\,=\,0
      \label{eq:xcr}
\end{equation}
and denote
\begin{equation}
      \Delta_{c}\,=\,{1 \over 2\ii\pi}\,F'_{0}(x_{c})\,=\,{1 \over
      2\pi}\,\Im(F'_{0}(x_{c}))
      \label{eq:delc}
\end{equation}
We expand  $F_{0}$ around some $x_{k}$ defined by \eq {eq:xkdef} and such that
\begin{equation}
      x_{k}-x_{c}\,=\, {\cal O}(N^{-1})
      \label{eq:xkco}
\end{equation}
and we get for the total free energy (by exactly the same calculation as in
sect.\ref{subsubsec.2.2.3})
\begin{eqnarray}
      \begin{array}{rcl}
      F\ &=\ &N^{2}\,F_{0}({x}_{k})\,-\,
      \ln\left({\theta_{3}(N{x_k})}\right)\\
      &&\vphantom{o}\\
      &&\, +\, F_1({x}_c) \,+\, {1\over 2}\,\ln\left({2\pi
      F''_{0}(x_{c})}\right)\,+\,{\cal O}(N^{-2})
      \end{array}
      \label{eq:F2cC}
\end{eqnarray}
where $\theta_{3}$ is the theta function with modular parameter
\begin{equation}
      \tau\,=\,{2\ii\pi\over F''_{0}(x_{c})}
      \label{eq:tauC}
\end{equation}
Only the
first two terms are important for calculating the two-point functions
and the orthogonal polynomials in the large $N$ limit.  This leading
term does not depend on $k$.  Indeed we have
\begin{equation}
      F'_{0}(x_{k})-F'_{0}(x_{c})\,=\, (x_{k}-x_{c}) F''_{0}(x_{c})
      \,+\,{\cal O}(N^{-1})
      \label{eq:FcFk}
\end{equation}
hence
\begin{equation}
     (x_{k}-x_{c})\,=\,{1\over N}\,\tau(k-N\Delta_{c}) \,+\,{\cal O}(N^{-2})
      \label{eq:xcxk}
\end{equation}
and using the periodicity relations of $\theta_{3}$ we can rewrite the
leading term for the free energy as
\begin{equation}
      N^{2}\,F_{0}({x}_{k})\,-\,
      \ln\left({\theta_{3}(N{x_k})}\right)
      \,=\,
      N^{2}\,F_{0}(n_{c}/N)\,-\,{\ii\pi\over\tau}u_{c}^{2}\,-\,
      \ln\left({\theta_{3}(u_{c})}\right)
      \label{eq:F2cC2}
\end{equation}
with
\begin{equation}
      n_{c}\,=\,{\mathrm{E}}[Nx_{c}]\quad,\quad
      u_{c}\,=\,[Nx_{c}]-\tau\,[N\Delta_{c}]
      \label{eq:InFr}
\end{equation}
where ${\mathrm{E}}[u]$ is the integer part of $u$ (largest integer
smaller than $u$) and $[u]=u-{\mathrm{E}}[u]$ is the fractional part of
$u$.  This does not depend on $k$ up to negligible terms of order
${\cal O}(N^{-1})$
(provided that condition \ref{eq:xkco} for $k$ holds).

One can now repeat the calculation of sect.\ref{sec.3} for the
2-point function.  Nothing is changed but we simply have to replace
$x_{c}$ by $x_{k}$ in the intermediate steps and to use \eq{eq:xcxk}
at the end of the calculation.
This amounts to replace the $x_{c}$ in the elliptic function $\sn^{2}$
by $x_{c}-\tau\Delta_{c}$. The final result for the two-point resolvent is
\begin{eqnarray}
      \omega_{}^{c}(\l,\mu) \, & = &
      \,-\,{1\over 4(\l-\mu)^{2}}\,
      \left[\left(1-\sqrt{{(\l-a)(\l-b)(\mu-c)(\mu-d)\over
      (\mu-a)(\mu-b)(\l-c)(\l-d)}}\right)+
      \left(\l\leftrightarrow\mu\right)\right]\nonumber\\
           & & -\,{(c-a)(d-b)\over 4 \ssq{\l}\ssq{\mu}}\
           \sn^{2}(N(x_{c}-\tau\Delta_{c})+{\scriptstyle {1\over 2}})
       \label{eq:f2ptC}
   \end{eqnarray}

Similar results holds for the orthogonal polynomials.
We simply have to consider the end-points $a,b,c,d$ for the mean-field
real parameter $x_{c}$ and to make the replacement
\begin{equation}
      Nx_{c}\ \to\ N(x_{c}+\tau\Delta_{c})
      \label{eq:xcdelc}
\end{equation}
in the elliptic functions involving $Nx_{c}$.  In any case these terms
depend only on the fractional parts of $Nx_{c}$ and of $N\Delta_{c}$.

A final interesting remark
on the periodicity properties of the non-universal term
$\sn^{2}(N(x_{c}+\tau\Delta_{c}))$ can
be made.  From the definition \ref{eq:delc} $\Delta_{c}$ corresponds
to a ``phase shift'' between the two arcs where the density of e.v. is
non zero.
   \begin{equation}
       \Delta_{c}\,=\,{1 \over 
2\ii\pi}\,F'_{0}(x_{c})\,=\,{\Gamma_{1}-\Gamma_{2}
       \over 2\ii\pi}
       \label{eq:DPhase}
   \end{equation}
where $N\Gamma_{\alpha}$ is the (constant) effective potential on the
arc $\alpha$.
The two periods $1$ and $\tau$ of the $\sn^{2}$ function correspond
respectively in term of eigenvalues to (i) transfer a single e.v. from
the first arc to the second one ($\delta Nx=\pm 1$), (ii) or to shift
the phase between the two arcs by $2\pi$ ($\delta N\Delta_{c}=\pm 1$).

\appendix{Multicut Case}
\label{manycuts}

Consider now a support of eigenvalues split into $s$ intervals:
\begin{equation}
  {\cal C}={\cal C}_1 \cup \dots \cup {\cal C}_s
\end{equation}

Let $n_i$ be the number of eigenvalues in each ${\cal C}_i$, and
$x_i=n_i/N$ the occupation ratio, which we denote collectively as a
vector: \begin{equation} x_i=\int_{{\cal C}_i}\D\l \rho(\l) \virg \vec{x} =
(x_1,\dots , x_{s-1}) \end{equation} Note that only $s-1$ of them are
independent since $x_1+\dots+x_s =1$.

As in the two-cut case (eq. \ref{eq:Fnexp}), the free energy at fixed
$\vec{n}$
admits a topological large $N$ expansion:
\begin{equation}
  F[V;\vec{n}] = N^2 F_0[V;\vec{x}] + N^0 F_1[V;\vec{x}] +{\mathcal{O}}(1/N^2)
\end{equation}
and as in \ref{eq:Z2csub}, the partition function can be written as a sum over
$\vec{n}$:
\begin{equation}
  Z= \ee{- F} = \sum_{\vec{n}} \, \ee{- F[V,\vec{n}] }
\end{equation}
The sum is dominated by the vicinity of the extremum $\vec{x}_c$ of
$F_0[V;\vec{x}]$:
\begin{equation}
  Z \sim \sum_{\vec{n}} \, \ee{-N^2 \left(F_0[V,\vec{x_{c}}] + i\pi
({\vec{n}\over N}-\vec{x}_{c}).\tau^{-1} ({\vec{n}\over N}-\vec{x}_{c})
  \right)
}
\where \left. {{\d}\over {\d \vec{x}}}F_0(\vec{x})\right|_{\vec{x}=\vec{x}_c}
  = \vec{0}
\end{equation}
$\tau$ is the $s-1\times s-1$ matrix defined by:
\begin{equation}
  \tau^{-1}_{ij} = {1\over 2i\pi}\left. {\d^2 F_0\over \d x_i \d x_j}
\right|_{\vec{x}=\vec{x}_c}
\end{equation}
Then, the summation over $\vec{n}$ yields:
\begin{equation}
  Z \sim \ee{-N^2 F_0[V,\vec{x}_{c}]} \,\,\theta (N\vec{x}_{c}|\tau)
\end{equation}
where $\theta(\vec{u}|\tau)$ is Riemann's theta function
\cite{Hyperelliptic} in genus $s-1$:
\begin{equation}
  \theta(\vec{u}|\tau) = \theta(\vec{u}) =
\sum_{\vec{n}} \ee{-i\pi (\vec{n}-\vec{u}).\tau^{-1}(\vec{n}-\vec{u})}
= \sum_{\vec{n}} \ee{i\pi \vec{n}.\tau \vec{n} } \ee{-2i\pi \vec{n}.\vec{u}}
\end{equation}
where $\tau$ is a $s-1\times s-1$ matrix, $\vec{u}$ is a $s-1$ 
component vector,
and $\vec{n}$ is a vector with integer coordinates.

The $\theta$ function obeys the relations ($\vec{k}$ being an arbitrary
integer vector):
\begin{equation}
  \theta(\vec{u}+\vec{k})=\theta(\vec{u})
   \virg   \theta(\vec{u}+\tau\vec{k}) = \ee{-i\pi(2\vec{u}.\vec{k} +
\vec{k}.\tau\vec{k})} \theta(\vec{u})
  \virg    \theta(-\vec{u})=\theta(\vec{u})
\end{equation}

Eventually the free energy at leading orders in $N$ is:
\begin{equation}\label{eq:Fmulticut}
  F \sim N^2 \left[ F_0[V,\vec{x}_{c}] - {1\over N^2}
  \ln{\theta(N\vec{x_c}|\tau)} +{1\over N^2}
F_1(\vec{x}_c) +\dots \right]
\end{equation}

It is now straightforward but lengthy to rederive the 2-point
correlation function and  the orthogonal polynomials from
\ref{eq:Fmulticut}.  One needs to differentiate \ref{eq:Fmulticut}
with respect to variations of the potential as in \ref{eq:2ptdeps} or
\ref{eq:PndS}, and express the hyperelliptical functions involved in
the calculation through prime forms (hyperelliptical generalization of
the $\theta_1$ function) \cite{Hyperelliptic}.  One should thus obtain
expressions similar to those of \cite{deift}.

\vfill\eject

\end{document}